\documentclass[onecolumn,superscriptaddress,showpacs,preprintnumbers,amsmath,amssymb]{revtex4-1}

\usepackage{soul}
\usepackage{amsmath}
\usepackage{amssymb}
\usepackage{graphicx}
\usepackage{morefloats}
\usepackage[usenames,dvipsnames]{xcolor}
\usepackage{blkarray}
\usepackage{verbatim}
\usepackage{hyperref}
\usepackage{subfigure}
\usepackage{natbib}
\usepackage{wrapfig}

\numberwithin{equation}{section}
\hypersetup{colorlinks=true, citecolor=orange, urlcolor=cyan, linkcolor=blue}

\begin{document}

\title{Colombian export capabilities: building the firms-products network}
\author{Matteo Bruno}
\affiliation{IMT School for Advanced Studies, P.zza S. Francesco 19, 55100 Lucca (Italy)}
\author{Fabio Saracco}
\affiliation{IMT School for Advanced Studies, P.zza S. Francesco 19, 55100 Lucca (Italy)}
\author{Tiziano Squartini}
\affiliation{IMT School for Advanced Studies, P.zza S. Francesco 19, 55100 Lucca (Italy)}
\author{Marco Due\~nas}
\affiliation{Department of Economics, International Trade and Social Policy, Universidad de Bogot\'a Jorge Tadeo Lozano, Bogot\'a (Colombia)}

\date{\today}

\begin{abstract}
In this paper we analyse the bipartite Colombian firms-products network, throughout a period of five years, from 2010 to 2014. Our analysis depicts a strongly modular system, with several groups of firms specializing in the export of specific categories of products. These clusters have been detected by running the bipartite variant of the traditional modularity maximization, revealing a bi-modular structure. Interestingly, this finding is refined by applying a recently-proposed algorithm for projecting bipartite networks on the layer of interest and, then, running the Louvain algorithm on the resulting monopartite representations. 
Important structural differences emerge upon comparing the Colombian firms-products network with the World Trade Web, in particular, the bipartite representation of the latter is not characterized by a similar block-structure, as the modularity maximization fails in revealing (bipartite) nodes clusters. This points out that economic systems behave differently at different scales: while countries tend to diversify their production --potentially exporting a large number of different products-- firms specialize in exporting (substantially very limited) baskets of basically homogeneous products.
\end{abstract}
\keywords{Complex Networks \and Inference Methods \and Exponential Random Graphs \and Null Models \and Economic Systems}
\pacs{89.75.Fb; 02.50.Tt; 89.65.Gh}

\maketitle

\section{Introduction\label{introduction}}

Exporting activities of countries have remarkable signals of complexity. By tradition, the understanding of the international trade has been of interest to politicians and economists. More recently, with the surge of the complex networks theory, the understanding of international trade has been enriched, providing information about the structure of industries and how it relates with countries growth, income, and development \cite{Hidalgo2007,Caldarelli2012}. This paper provides new and inspiring evidence on the study of the productive capacity of a nation based on its exports. More precisely, we analyze the bipartite network of Colombian exports using data at the firm level for the period 2010-2014 by employing tools developed within the field of information theory and complex networks analysis \cite{Saracco2016b}. 

The bipartite network derives from considering the type of products exported by firms, ending up with two layers: firms and products. We are interested in the understanding of the projections on those layers, carefully dealing with the statistical significance of the similarity measure employed. Our approach is based on the maximization of the constrained Shannon entropy with the available information about the system at hand: since we are interested in detecting common patterns of economic activities characterizing Colombian firms, we employ properly-defined constraints to define benchmarks for testing the statistical significance of the observed patterns. As a result, the building blocks in our analysis rely on connecting nodes by their similarity in both layers: firms are connected because they export a significantly large number of common products and products are connected because they are jointly exported by a significantly large set of firms.

Diversification of products is related to the growth of firms by the expansion into new activities or markets. It has been argued that firms accumulate specific capabilities that can be used to produce different products or to enter different industries \cite{penrose1959}. From the perspective of cost minimizing firms, economies of scope are revealed when the cost of joint production of different products is less than the cost of producing all of them separately \cite{panzar_willing_81}. In this sense, considering that firms' capabilities are revealed in production intensity and product portfolio, the focus has been on the understanding on why firms diversify, how diversification emerges, and its relation to firms' performance \cite{teece1980, teece1982, teece1994}.

At the level of international trade, the network analysis made possible to build taxonomies of products and countries that allowed a better understanding of countries' exporting capabilities \cite{Hausmann2007,Tacchella2012,Cristelli2013}. Challenging the Ricardo's fundamental comparative advantage theory, the surprising outcome was that the bipartite matrix was triangular (sorting by countries capabilities) rather than block-diagonal\footnote{The latter is the case in which comparative advantages of countries induce specialization in a few products according to their factor and technological endowments.}. Nevertheless, developed countries have the capabilities to produce and export a wide variety of sophisticated and unsophisticated products, while the developing countries reveal much more restricted capacities that are related to the exports of less complex products.

There are important differences at comparing the results of the present paper with those in international trade. A firm is certainly much more constrained in its production, both in scale (volumes) and scope (number of different products). Therefore, the expected outcome of the present study must be by construction different. Obviously, we attempt to go deeply in the well-studied complexities of the economic systems giving important insights of the fitness of countries. And more precisely, we aim to provide inputs to understand the process of firms product diversification as a consequence of combinations of (unobserved) capabilities. In this sense, we could assume that the higher the capabilities of a firm, the higher the number of the exported products. Thus, we attempt to discover if also at the country level firms and products get together in meaningful communities, and also we aim to recognize the differences and similarities between the micro and macro levels: firms versus countries.

The paper is organized as follows. In Section \ref{data} we describe the dataset and the data cleaning process. In Section \ref{methods} we provide a detailed explanation of the methods employed for the present analysis. In Section \ref{results} we illustrate the results of our study of the Colombian firms-products network and compare them with the corresponding ones, observed in the World Trade Web. Finally, we conclude with a general discussion of our results.

\section{Data\label{data}}

\paragraph*{Colombian export data.} We study the Colombian exports as a bipartite, undirected, binary network: firms and products constitute the nodes of the two different layers and intra-layer links are not permitted. For the description of the Colombian firms-products (CFP in what follows) network, we use all export transactions of manufacturing products reported at the Colombian Customs Office (Direcci\'on de Impuestos y Aduanas Nacionales, DIAN) and collected by the Colombian Bureau of Statistics (Departamento Administrativo Nacional de Estad\'istica, DANE), for the period  2010 and 2014. We removed all transactions related to re-exports of products elaborated in other countries. Each shipment has a unique seller ID, which we use as the firm identifier, the date, a 6-digit harmonized system (hs) characterizing the product, the destination and the US dollar value of the transaction. For the 2010-2011 period, products are described by hs2007 coding while, in the following period, by hs2012 coding (6-digits in both cases). For the comparison, we used the conversion table provide by the UN~\footnote{\href{https://unstats.un.org/unsd/trade/classifications/correspondence-tables.asp}{https://unstats.un.org/unsd/trade/classifications/correspondence-tables.asp}}.\\

\paragraph*{World Trade Web data.} Data concerning the World Trade Web (WTW) are represented as a bipartite, undirected, binary network as well: countries and products constitute the nodes of the two different layers and intra-layer links are not permitted. For our analyses we use the BACI World Trade Database \cite{BACI2013}. Products are described by hs2007 coding at 4 digits.\\

\paragraph*{Data cleaning procedure.} We filter out small firms in the CFP and small countries in the WTW, since they would bring very little information. In the CFP, we removed firms with a total yearly export volume lower than current $10^4$ USD (results do not change upon varying such a threshold). In addition, a \emph{Revealed Comparative Advantage (RCA)}~\cite{Balassa1977} threshold is applied. This procedure, which is standard in the analysis of international trade, consists in comparing, for every firm/country, the share of each export product value with the global (i.e. over the entire dataset) analogous. In formulas, if $w_{ip}$ is the value of the export of the firm/country $i$ for the product $p$, the RCA reads:
\begin{equation}
RCA_{ip}=\dfrac{\dfrac{w_{ip}}{\sum_{i'}w_{i'p}}}{\dfrac{\sum_{p'}w_{ip'}}{\sum_{i',p'}w_{i'p'}}}=\dfrac{w_{ip}}{s_it_p/W}
\end{equation}
where we have defined the strength of firm/country $i$ as $s_i=\sum_{p'}w_{ip'}$, the strength of product $p$ as $t_p=\sum_{i'}w_{i'p}$ and $W=\sum_{i',p'}w_{i'p'}$ is the total, exported volume. If the firm/country share is greater than the global share, i.e. if $RCA\ge1$ (or, equivalently, $w_{ip}\ge\dfrac{s_it_p}{W}$) its ``exporting performance'' on the given commodity is interpreted as being above-average and the entry is validated.

The output of this cleaning procedure is a rectangular binary matrix $\mathbf{M}$ (i.e. the biadjacency matrix of our bipartite, undirected, binary network). In the case of the CFP biadjacency matrix $\mathbf{M}_{CFP}$, the number of firms and products will be indicated, respectively, as $F$ and $P$. For simplicity, we omitted temporal subscripts, but in the case of the CFP both $F$ and $P$ vary over time. The matrix generic entry $m_{ip}$ is 1 if firm/country $i$ exports an amount of product $p$ above the RCA threshold; otherwise, $m_{ip}=0$. Each row represents the basket of products of a given firm/country and, similarly, each column represents the set of exporters of a given product. 

\begin{figure}[t!]
\includegraphics[width=.45\textwidth]{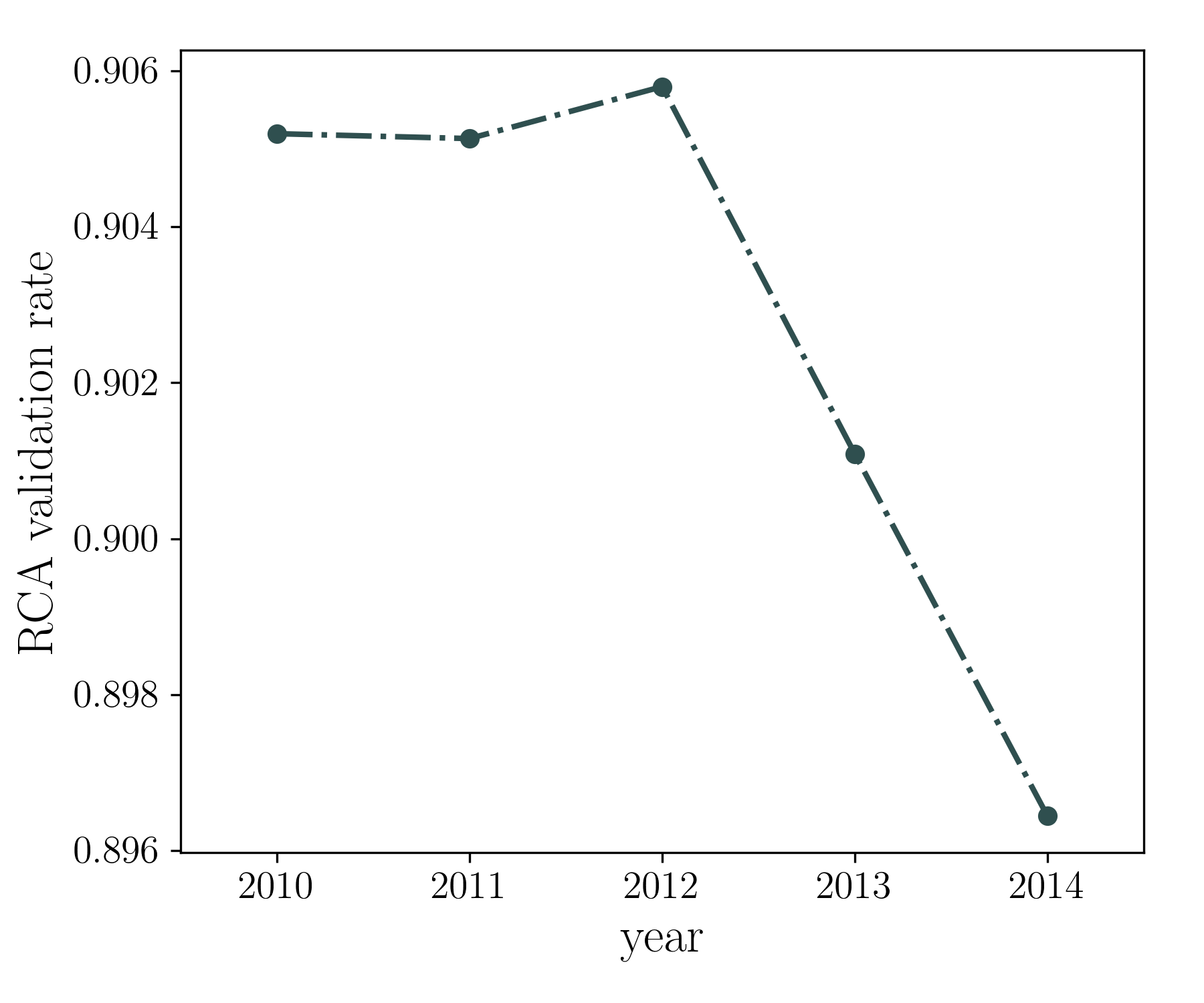}
\includegraphics[width=.45\textwidth]{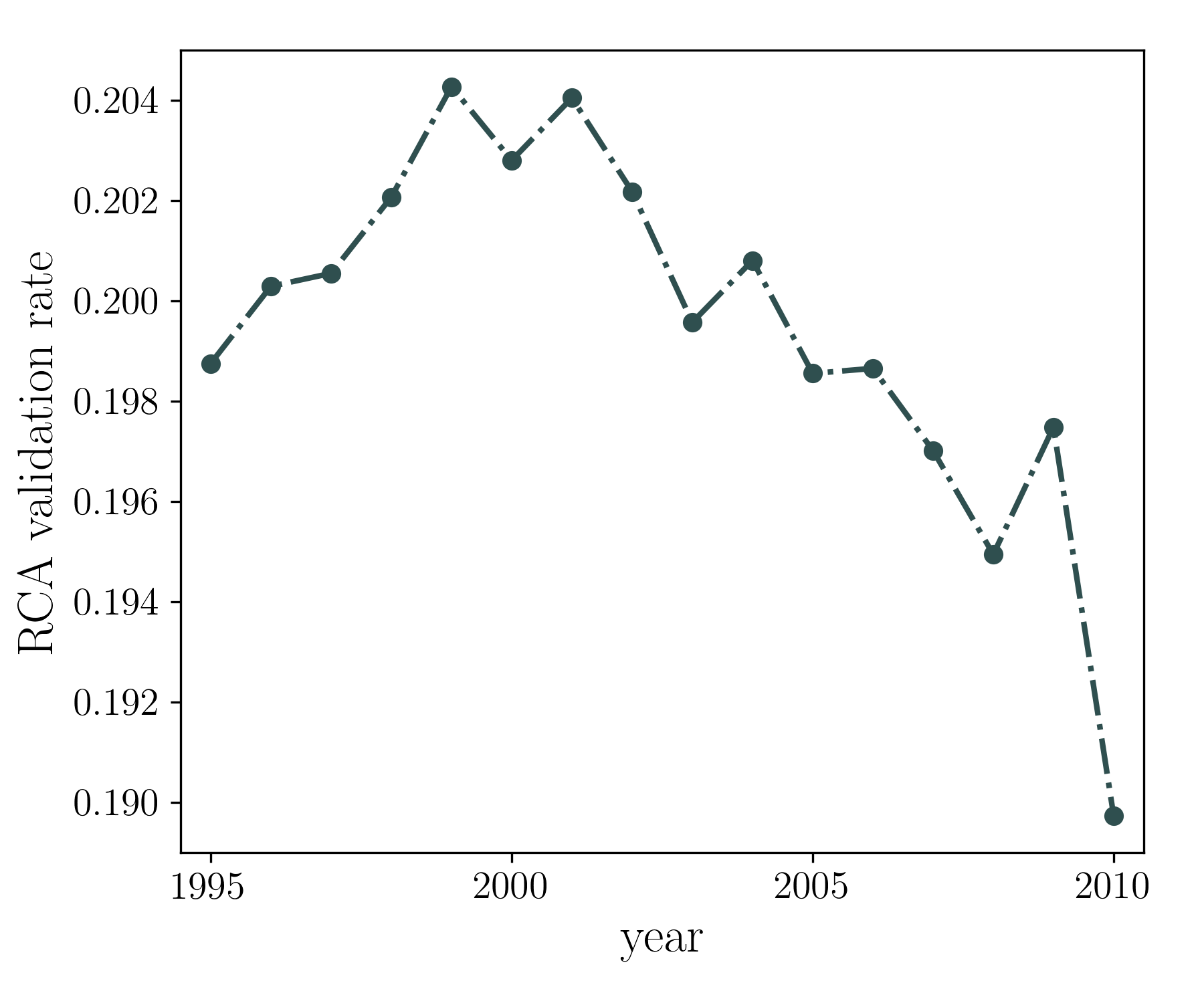}
\caption{Percentage of validated links by the RCA on the Colombian export dataset (left) and the WTW (right). While the RCA behaves as a selective filter on the World Trade Web, it is much less effective on the network of Colombian exports.}
\label{figrca}
\end{figure}

Interestingly, the obtained bipartite networks have different connectances depending on the system analyzed: the WTW density of links ranges from 0.09 to 0.13, while in the Colombian dataset its order of magnitude is steadily $\simeq 10^{-3}$. The percentage of links validated by the RCA thresholding procedure differs for the Colombian national exports and for the WTW, passing from $\simeq0.9$ for the CFP to $\simeq0.2$ for the WTW (see also fig. \ref{figrca}). This implies that there is no such a big difference between the topological structure of the Colombian export network and of its binarized version obtained by employing the RCA threshold. In turn, this means that there is relatively low competition from national producers in the products exported by firms.

As a general observation, while the number of products of the CFP network remains practically constant throughout the considered temporal period ($P\simeq 3100$), the number of firms keeps increasing, moving from $F\simeq 4800$ in 2010 to $F\simeq 5100$ in 2014. Moreover, while the number of persistent firms (i.e. the ones that are present throughout the entire time period) is $\simeq 2240$, the number of persistent products is $\lesssim 2400$, i.e. a high percentage of the total. For the sake of the comparison, the WTW is characterized by a number of countries $\simeq 140$ and a number of products $\simeq 1131$ throughout all years.


\section{Methods\label{methods}} 

In order to analyse the Colombian firms-products network we apply a recently-proposed algorithm to obtain monopartite representations of bipartite networks. Since the method is perfectly general, in what follows we will index the nodes corresponding to the rows of the biadjacency matrix with the letter $r=1\dots R$ and those corresponding to the columns of it with the letter $c=1\dots C$. In our specific case, firms lie along the rows and products lie along the columns.

As in \cite{Saracco2016b}, we implement the following four-steps recipe, prescribing to \emph{a)} choose a specific pair of nodes belonging to the layer of interest, say $r$ and $r'$, and measure their similarity, \emph{b)} quantify the statistical significance of the measured similarity with respect to a properly-defined null model, \emph{c)} link nodes $r$ and $r'$ if, and only if, such a similarity is found to be significant, \emph{d)} repeat the steps above for every pair of nodes.\\

\paragraph*{Measuring nodes similarity.} The most straightforward approach to quantify nodes similarity is counting the number of common neighbors $V_{rr'}$ shared by nodes $r$ and $r'$ and defined as:

\begin{equation}\label{eq:V}
V_{rr'}=\sum_{c}m_{rc}m_{r'c}=\sum_{c}V_{rr'}^c.
\end{equation}
In (\ref{eq:V}) we adopted the definition $V_{rr'}^c\equiv m_{rc}m_{r'c}$ for the basic quantity of our approach, satisfying the relationship $V_{rr'}^c=1$ if, and only if, both $r$ and $r'$ share the (common) neighbor $c$. In other words, we quantify the similarity between any two Colombian firms (products) by counting the number of their co-exported products (co-exporting firms). As discussed in \cite{teece1994,Bottazzi2010}, alternative measures for quantifying similarity - also known with the name of \emph{relatedness} - indeed exist. Our choice has been dictated by the intuitive meaning of the number of co-occurrences, beside its analytical tractability under the null hypothesis described below.\\

\paragraph*{Quantifying the significance of nodes similarity.} The second step of our algorithm prescribes to quantify the statistical significance of the similarity of $r$ and $r'$. To this aim, a benchmark is needed: a natural choice leads to adopt the Exponential Random Graph (ERG) class of null-models \cite{Park2004,Garlaschelli2008, Squartini2011,Fronczak2012,Saracco2015}. Briefly speaking, the ERG formalism rests upon the constrained maximization of Shannon entropy, a procedure aimed at maximizing the uncertainty about the system at hand except for what is known about it (represented by the constraints). Such a recipe ensures that a maximally unbiased inference is carried out, thus minimizing the risk of drawing unsupported conclusions about our data.

More quantitatively, maximizing $S=-\sum_{\mathbf{M}}P(\mathbf{M})\ln P(\mathbf{M})$ under the chosen constraints leads to assign to the generic bipartite network $\mathbf{M}$ an exponential probability $P(\mathbf{M})=\frac{e^{-H(\vec{\theta},\:\vec{C}(\mathbf{M}))}}{Z(\vec{\theta})}$, whence the name of the formalism. The numerical value of $P(\mathbf{M})$ is, then, determined by the vector $\vec{C}(\mathbf{M})$ of topological constraints \cite{Park2004}. In order to determine the unknown parameters $\vec{\theta}$, the likelihood-maximization recipe can be adopted: given an observed biadjacency matrix $\mathbf{M}^*$, it translates into solving the system of equations $\langle \vec{C}\rangle(\vec{\theta})=\sum_{\mathbf{M}}P(\mathbf{M})\vec{C}(\mathbf{M})=\vec{C}(\mathbf{M}^*)$, which prescribes to equate the ensemble averages $\langle \vec{C}\rangle(\vec{\theta})$ to their observed counterparts, $\vec{C}(\mathbf{M}^*)$ \cite{Squartini2011,Garlaschelli2008}. The null model we have considered in the present paper is known as the Bipartite Configuration Model (BiCM) \cite{Saracco2015} and is defined by constraining the degrees of nodes belonging to both layers. Enforcing the nodes degrees allows us to write $P(\mathbf{M})$ in a factorized form, i.e. as the product of pair-specific probability coefficients:

\begin{equation}
P(\mathbf{M})=\prod_{r=1}^R\prod_{c=1}^Cp_{rc}^{m_{rc}}(1-p_{rc})^{1-m_{rc}}=\prod_{r=1}^Rx_r^{d_r(\mathbf{M})}\prod_{c=1}^Cy_c^{u_c(\mathbf{M})}\prod_{r=1}^R\prod_{c=1}^C(1+x_ry_c)^{-1}
\end{equation}
depending on the firms' degree $d_r,\:r=1\dots R$ and on the products degree $u_c,\:c=1\dots C$ (whose names, \emph{diversification} and \emph{ubiquity}, we mutuate from the country-specific analogue \cite{Tacchella2012}) $x_r$ and $y_c$ being the Lagrange multipliers associated to the constrained degrees. Under the BiCM, the probability $p_{rc}$ that a link exists between firm $r$ and product $c$) reads:

\begin{equation}
p_{rc}=\frac{x_ry_c}{1+x_ry_c}
\label{prob}
\end{equation}
its numerical value being determined by the likelihood-maximization conditions:

\begin{equation}
\left\{ \begin{array}{ll}
\langle d_r\rangle&=\sum_r p_{rc} = d_r^*,\:r=1\dots F\\
\langle u_c\rangle&=\sum_c p_{rc} = u_c^*,\:c=1\dots P
\end{array}
\right.\\
\label{sys}
\end{equation}
prescribing that the average values of firms diversification and products ubiquities match the observed counterparts $\{\vec{d}^*\}$ and $\{\vec{u}^*\}$.

Since ERG models with linear constraints treat links as independent random variables, the presence of each $V_{rr'}^c$ can be regarded as the outcome of a Bernoulli trial:

\begin{eqnarray}
f_\text{Ber}(V_{rr'}^c=1)&=&p_{rc}p_{r'c},\\
f_\text{Ber}(V_{rr'}^c=0)&=&1-p_{rc}p_{r'c};
\end{eqnarray}
it follows that, once $r$ and $r'$ are chosen, the events describing the presence of the single $V_{rr'}^c$ patterns are independent random experiments: this, in turn, implies that each $V_{rr'}$ is a sum of independent Bernoulli trials, each one described by a different probability coefficient. The distribution describing the behavior of each $V_{rr'}$ turns out to be the so-called Poisson-Binomial \cite{Saracco2016b}. Measuring the statistical significance of the nodes similarity $r$ and $r'$ translates into calculating a p-value on the aforementioned Poisson-Binomial distribution, i.e. the probability of observing a number of patterns greater than, or equal to, the observed one (which will be indicated as $V_{rr'}^*$):

\begin{equation}
\text{p-value}(V_{rr'}^*)=\sum_{V_{rr'}\geq V_{rr'}^*}f_{\text{PB}}(V_{rr'}).
\end{equation}

Upon repeating such a procedure for each pair of nodes, we obtain a $R\times R$ matrix of p-values \cite{Saracco2016b,Hong2013}. Since this operation is computationally costly, for sparse networks it is possible to rest upon the approximation of the Poisson-Binomial distribution prescribing to substitute the variable $V_{rr'}$ with a Poisson variable having the same expected value $\mu$. The error of such an approximation is quantified by Le Cam's theorem \cite{Deheuvels1989,Volkova1996,Hong2013}:

\begin{equation}
\sum \limits_{k=0}^C\left\vert f_{PB}(V_{rr'}=k)-\frac{\mu^k \exp(-\mu)}{k!}\right\vert<2\sum_{c=1}^C(p_{rc}p_{r'c})^2.
\label{Poisson approximation}
\end{equation}

A comparison of the BiCM with alternative null models has been carried out in a similar-in-spirit analysis \cite{Saracco2016}.\\

\paragraph*{Validating the projection.} In order to understand which p-values are to be retained, a statistical procedure accounting for the fact that we are testing multiple hypotheses at a time is needed. In the present paper we apply the so-called False Discovery Rate (FDR) procedure \cite{Benjamini1995}. Whenever $M$ different hypotheses, $H_1\dots H_M$, characterized by $M$ different p-values, must be tested at a time, FDR prescribes to, first, sort the $M$ p-values in increasing order, $\text{p-value}_1\leq\dots\leq \text{p-value}_M$ and, then, to identify the largest integer $\hat{i}$ satisfying the condition

\begin{equation}
\text{p-value}_{\hat{i}}\leq\dfrac{\hat{i}t}{M}
\label{threshold}
\end{equation}
with $t$ representing the usual single-test significance level (e.g. $t=0.05$ or $t=0.01$). The third step of the FDR procedure prescribes to reject all the hypotheses whose p-values are less than, or equal to, $\text{p-value}_{\hat{i}}$ (i.e. all p-values satisfying $\text{p-value}_1\leq\dots\leq \text{p-value}_{\hat{i}}$). Notably, FDR allows one to control for the expected number of false ``discoveries'' (i.e. incorrectly-rejected null hypotheses), irrespectively of the independence of the hypotheses tested\footnote{Our hypotheses, for example, are not independent, since each observed link affects the similarity of several pairs of nodes}. In our case, the FDR prescription translates into adopting either the threshold $\hat{i}t/\binom{R}{2}$ or the threshold $\hat{i}t/\binom{C}{2}$ depending on the layer whose projection we are interested in. For the sake of clarity, each pair of nodes whose corresponding p-value passes the FDR validation is joined by a binary, undirected link in the corresponding projection. In what follows, we have used a single-test significance level of $t=0.05$.\\

\paragraph*{Testing the projection algorithm.} In order to test the performance of our method, the Louvain algorithm \cite{Blondel2008} has been run on the validated projections of the real networks considered for the present analysis. Since Louvain algorithm is known to be order-dependent \cite{Fortunato2010}, we considered a number of outcomes of the former equal to the size of the projected network - each one obtained by randomly reshuffling the order of nodes taken as input - and chose the one providing the maximum value of the modularity \cite{Saracco2016b,Fortunato2010}\\

\begin{figure}[t!]
\includegraphics[width=0.35\textwidth]{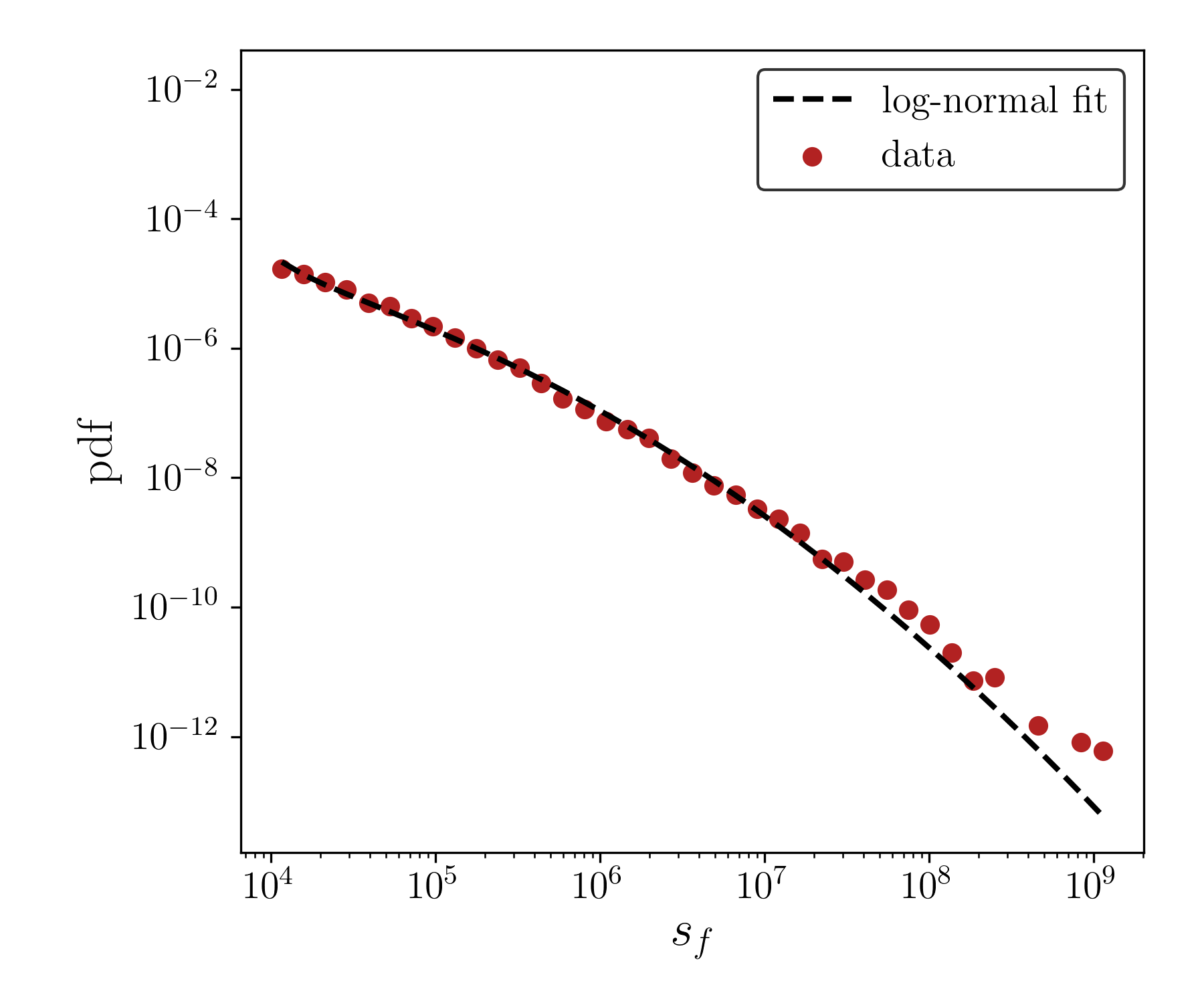}
\includegraphics[width=0.35\textwidth]{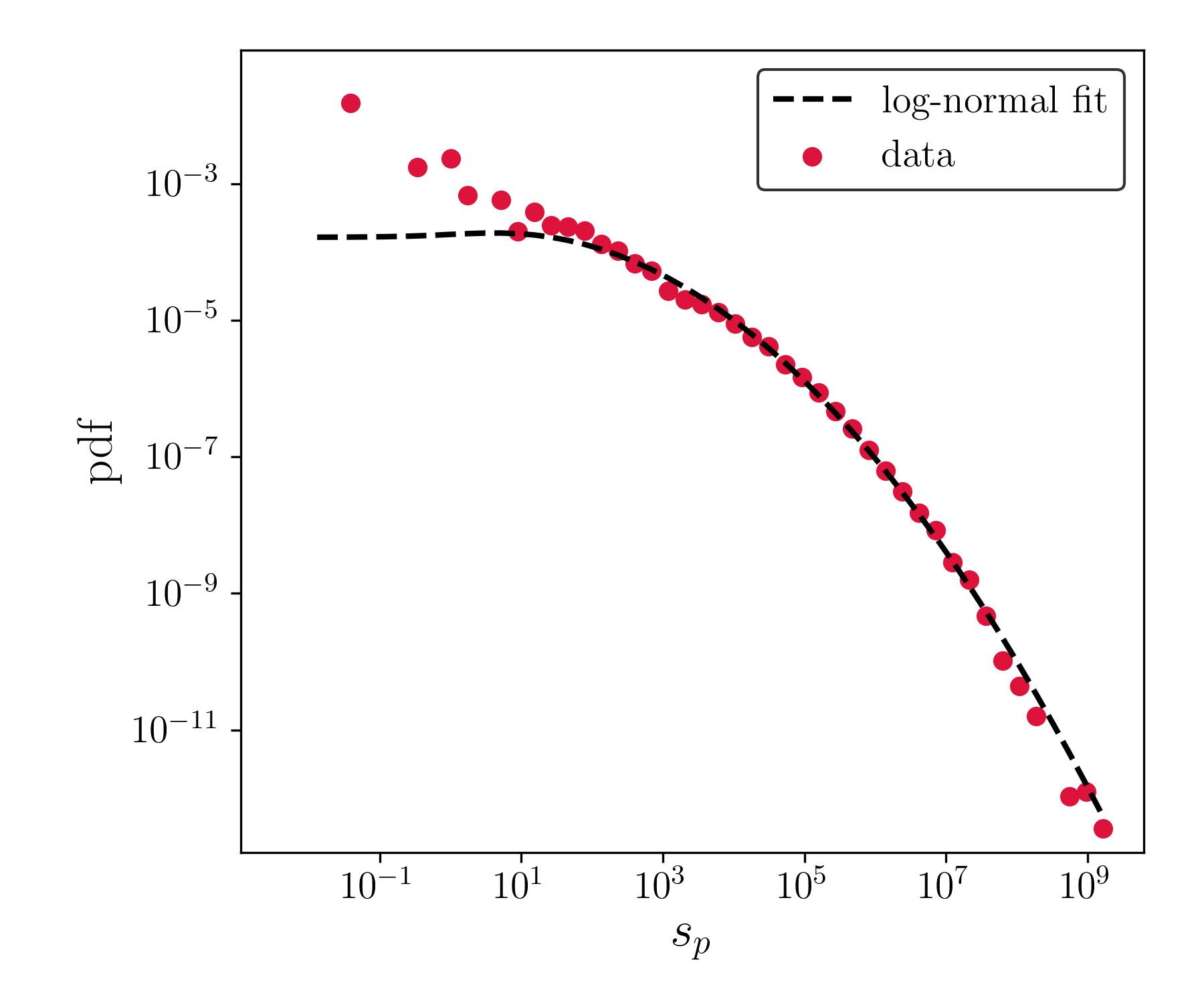}
\includegraphics[width=0.35\textwidth]{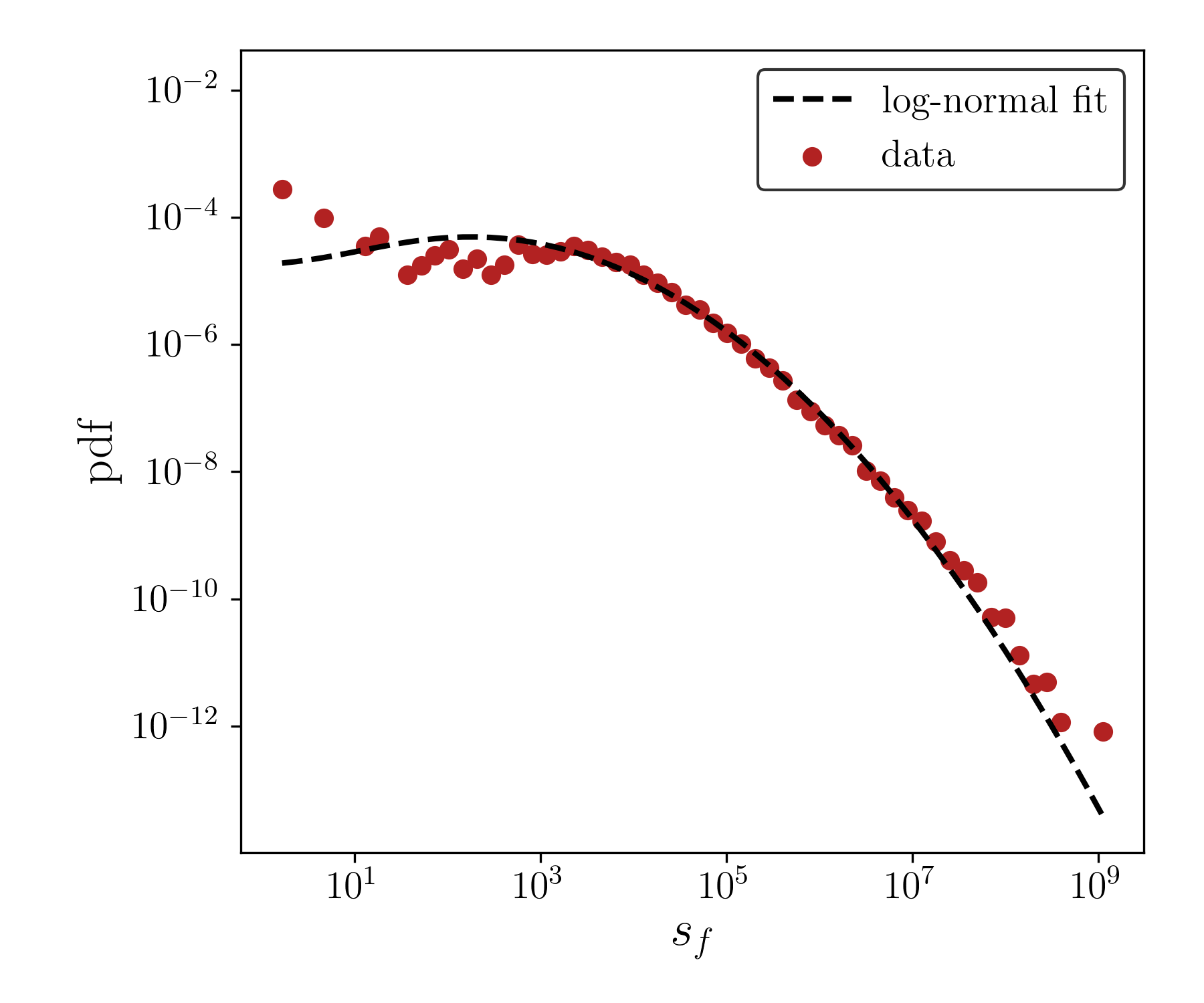}
\includegraphics[width=0.35\textwidth]{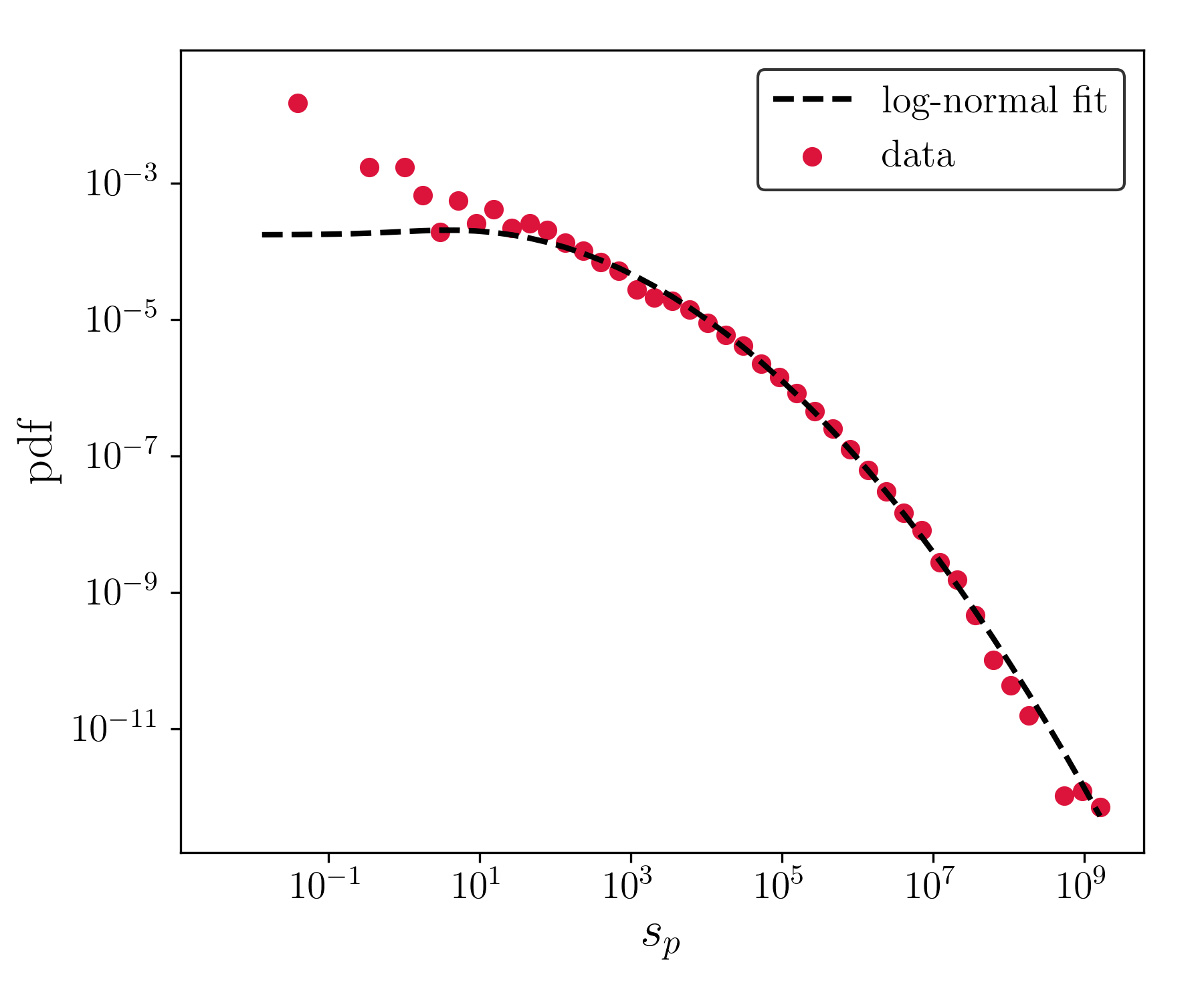}
\caption{Distributions of firms strengths (left) and products strengths (right) for the CFP network in 2010, before applying the threshold at $10^4$ USD (bottom panel) and after applying such a threshold (top panel): from a qualitative point of view, results do not change. A KS test does not reject the hypothesis that strengths are log-normally distributed.}
\label{figsdis}
\end{figure}

\paragraph*{Statistical analysis.} Our ensemble method allows the statistical significance of a large number of topological quantities to be tested. In order to quantify to what extent the considered null model is able to capture the real structure of the network, one can compare the observed and expected value of any quantity of interest $X$ via the so-called \emph{z-score}, defined as $z_X=\frac{X(\mathbf{M}^*)-\langle X\rangle}{\sigma_X}$ where $\langle X\rangle\simeq\sum_{\mathbf{M}}\tilde{P}(\mathbf{M})X(\mathbf{M})$ and $\sigma_X\simeq\sum_{\mathbf{M}}\tilde{P}(\mathbf{M})[X(\mathbf{M})-\langle X\rangle]^2$ are the sampling moments, computed according to the sampling frequencies $\tilde{P}(\mathbf{M})$. The latter are the sampling-induced approximations of the ensemble frequencies $P(\mathbf{M})$, computed by explicitly generating a sufficiently large number of network configurations (in our case, 1,000). Whenever the ensemble distribution of the quantity $X$ closely follows a Gaussian, $z$-scores can be attributed the usual meaning of standardized variables, enclosing the $99.7\%$ of the probability distribution within the range $z_X\in[-3,3]$: any discrepancy between observations and expectations leading to values $|z_X|>3$ can thus be interpreted as statistically significant.

However, when the ensemble distribution of the quantity $X$ deviates from a Gaussian, $z$-scores cannot be interpreted in the aforementioned way and an alternative procedure is needed: here, we have computed the \emph{box plots}. Box plots are intended to sum up a whole probability distribution by showing no more than five percentiles: the 25th percentile, the 50th percentile and the 75th percentile (usually drawn as three lines delimitating a central box), plus the $0.15$th and the 99.85th percentiles (usually drawn as whiskers lying at the opposite sides of the box). Box plots can, thus, be used to assess the statistical significance of the observed value of $X$ against the null value output by the BiCM.

\section{Results\label{results}}

The results of our analysis refer to the year 2010 for both the CFP network and the WTW. However, they are robust across years.

\subsection{Node degree and strength distributions}

Let us start by describing some empirical findings about the system under analysis, concerning the distribution of firms and products strengths. As fig. \ref{figsdis} shows, both distributions seem to be well fitted by a log-normal (as confirmed by a Kolmogorov-Smirnov (KS) test that does not reject such an hypothesis) in agreement with the the evidence for the monopartite WTW \cite{fagiolo2009pre} and also at the firm level \cite{Bee_etal_2017,Campi_etal_2017}.

\begin{figure}[t!]
\includegraphics[width=0.35\textwidth]{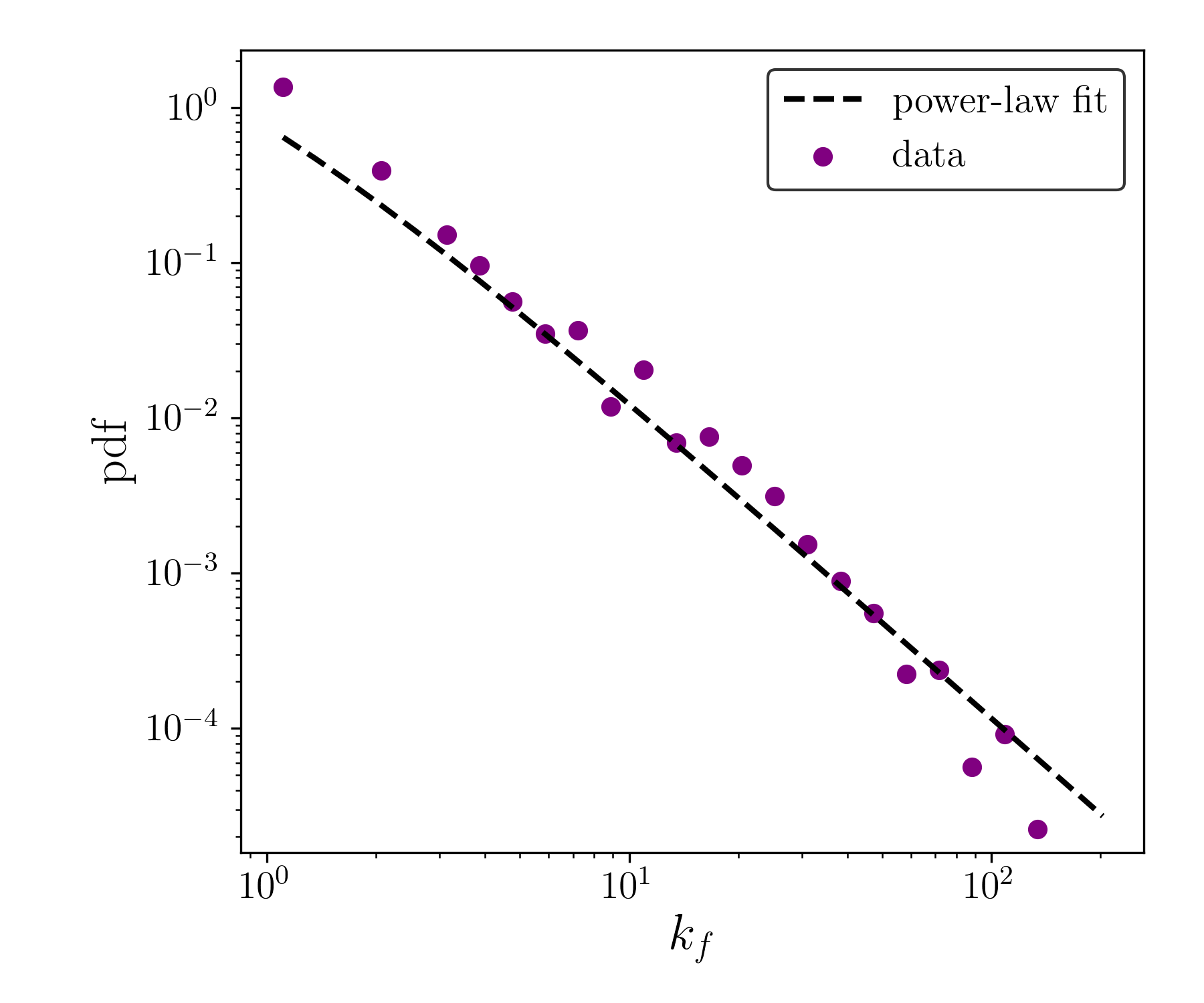}
\includegraphics[width=0.35\textwidth]{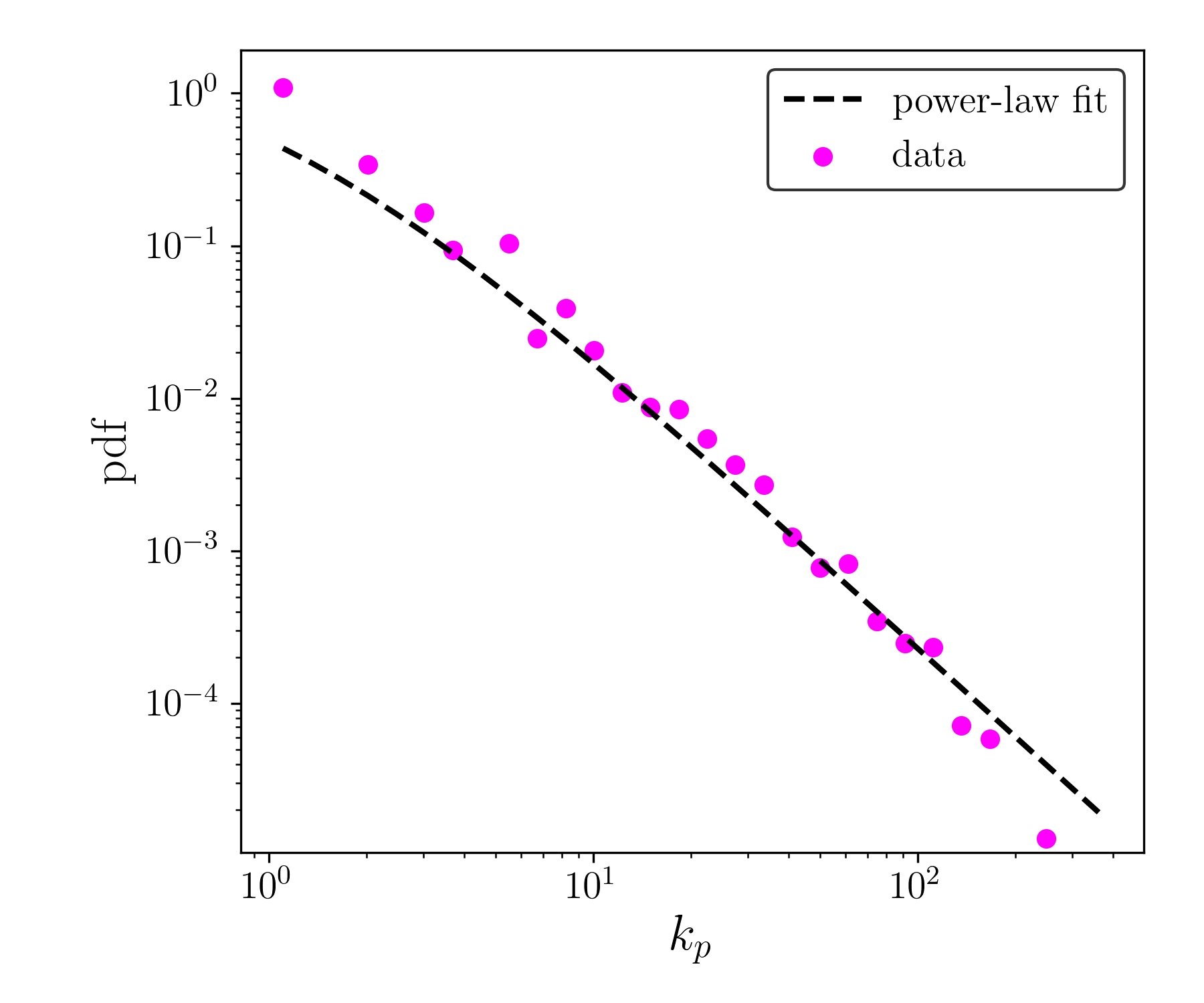}
\includegraphics[width=0.35\textwidth]{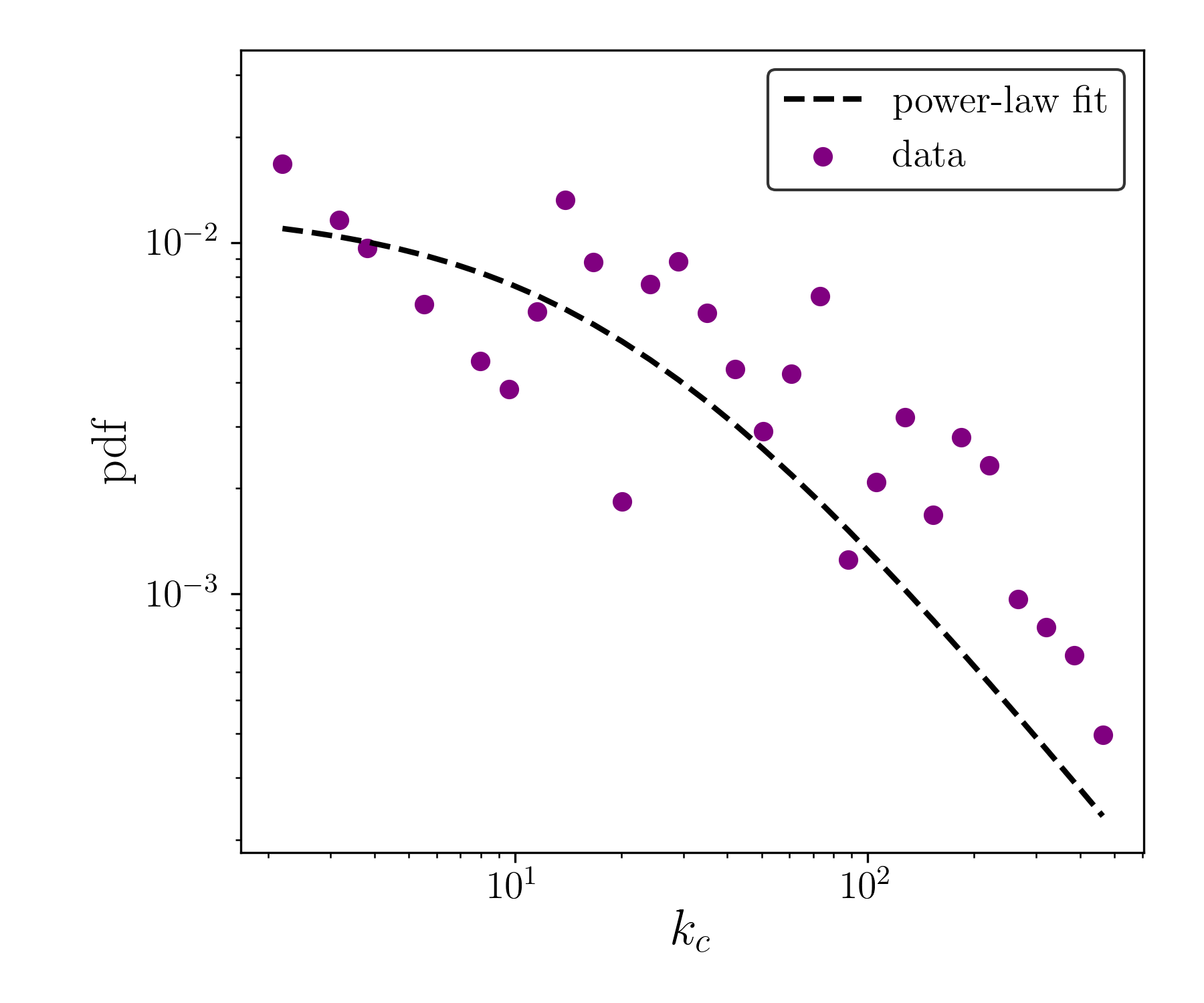}
\includegraphics[width=0.35\textwidth]{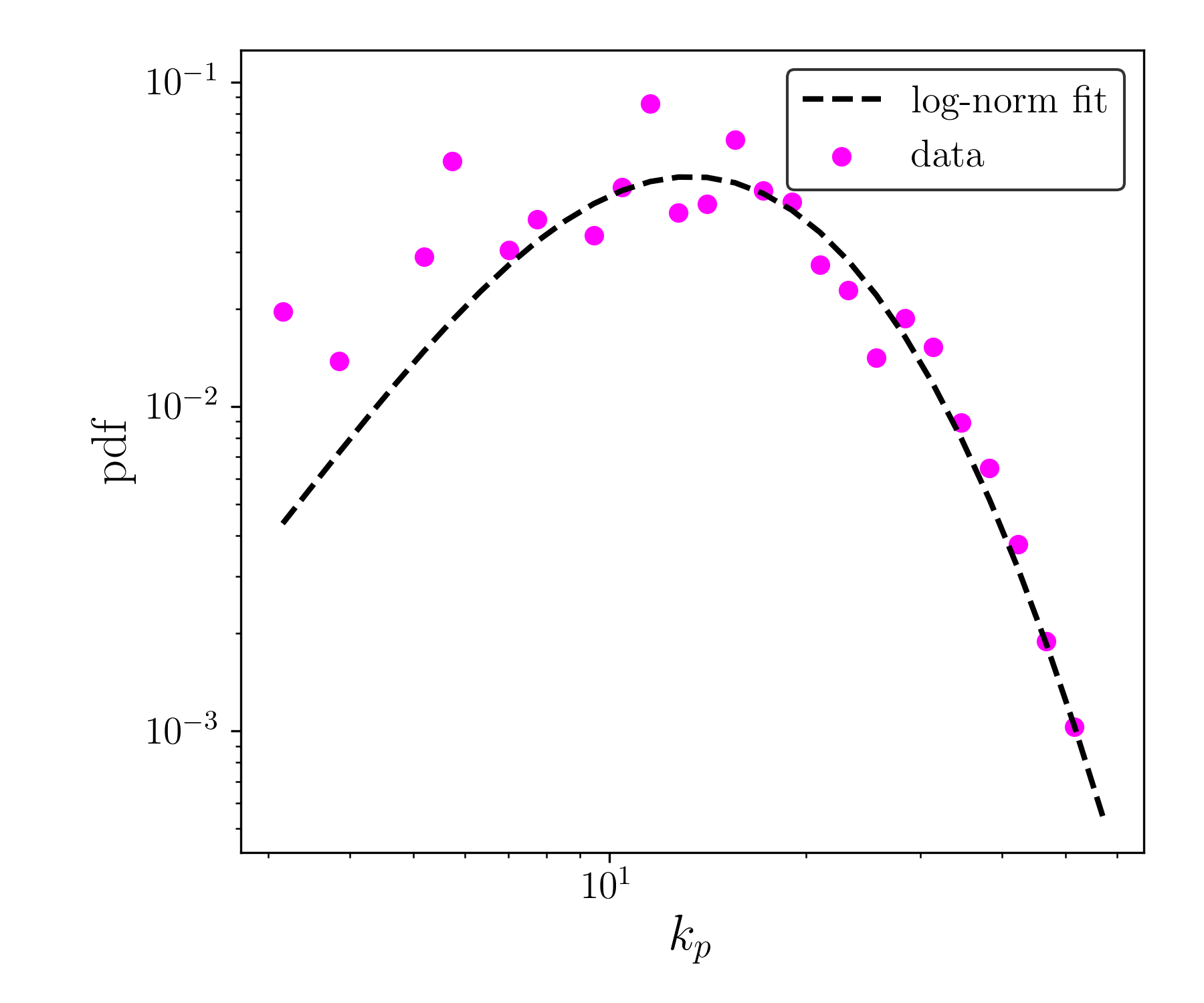}
\caption{Top panel: distributions of firms degrees (left column) and products degrees (right column) for the CFP network in 2010. Bottom panel: distributions of countries degrees (left column) and products degrees (right column) for the WTW in 2010. All distributions refer to the thresholded dataset. A KS test does not reject the hypotheses described in the insets.}
\label{figdis}
\end{figure}

Let us now move to the description of the degree distributions of firms and products. Heavy-tailed distributions for the values of the aforementioned quantities can be observed, across all considered years: the KS test does not reject the hypothesis that both distributions are compatible with power-laws (see the top panels of fig. \ref{figdis}). As a comparison, let us consider the distributions describing the countries and products degrees of the WTW: as the bottom panel of fig. \ref{figdis} shows, the distribution of the products degrees is compatible with a log-normal. On the other hand, although the distribution of the firms degrees appears to be a bit more noisy, a KS test does not reject the hypothesis that it follows a power-law.

\subsection{Nodes degrees and strengths correlation}

\begin{figure}[t!]
\includegraphics[width=0.9\textwidth]{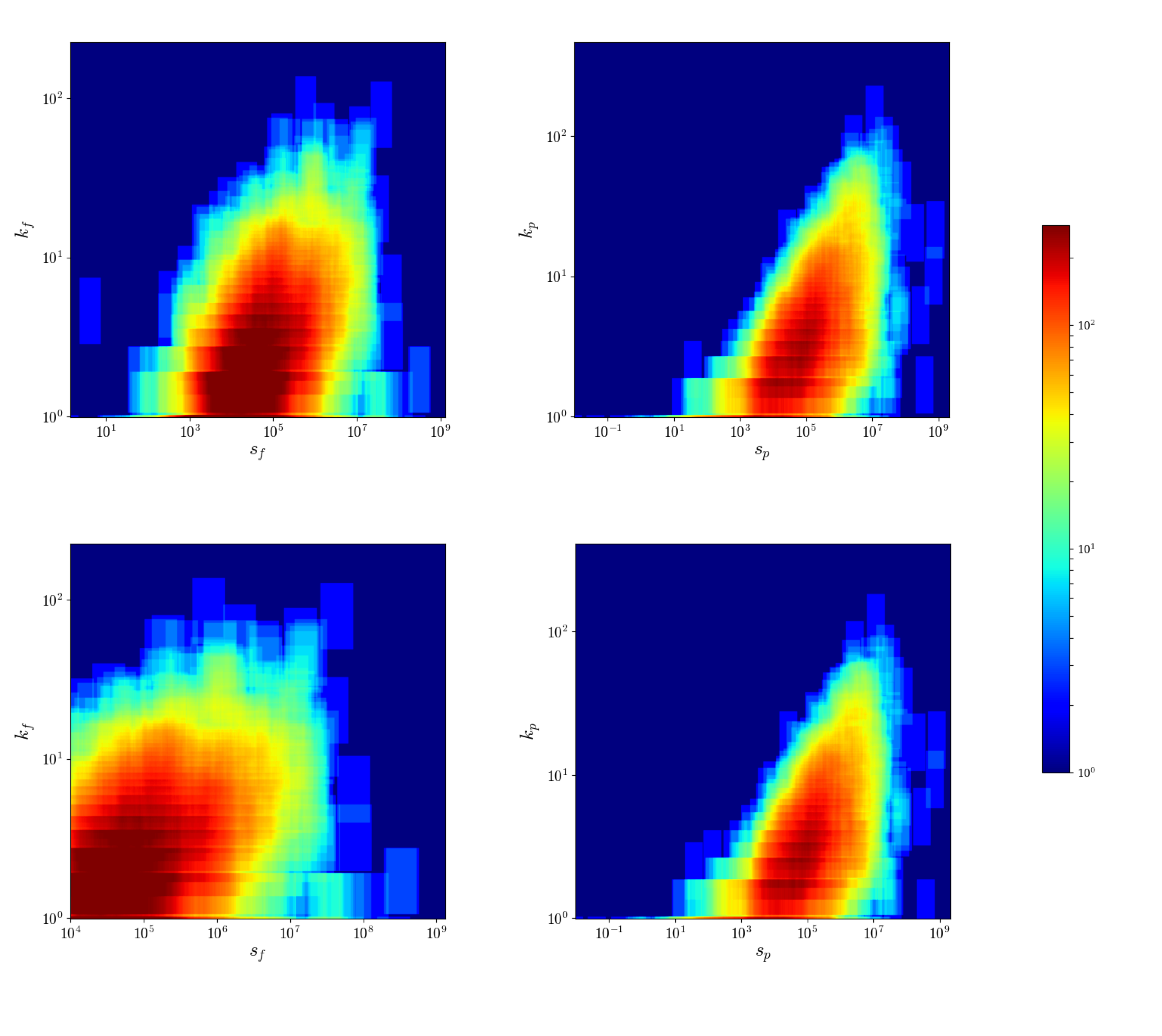}
\caption{Degree VS strength heatmap for the firms (left panel) and products (right panel) of the CFP network, after (top panel) and before (bottom panel) applying the threshold at $10^4$ USD. The heatmaps were obtained counting the number of points falling in sliding windows in log-log scale, and the color goes from blue to red as the density of points increases. The Spearman coefficient, computed on the thresholded dataset, is around 0.65 for products and 0.40 for firms. In the WTW case, it rises to 0.90 for countries and 0.73 for products.}
\label{figfit}
\end{figure}

The relationship between nodes degrees and strengths is found to be close related in several economic and financial networks, be they monopartite \cite{Cimini2015} or bipartite \cite{Squartini2017}. Briefly speaking, strengths are found to be positively correlated with the degrees, reflecting the fact that countries with a larger number of neighbors are also the ones exporting a larger volume of products. This seems to hold true also for the CFP network, as fig. \ref{figfit} shows. There is huge heterogeneity, especially for firms with medium to large strength. While it is expected that firms with relatively low total exports have a low scope, firms with higher total trade can be very specialized in a few products (even a single product) or highly diversified. As a consequence, this pattern is observed in the product projection, some products are intensively produced by a few firms and some are ubiquitous and produced intensively by many firms.

\subsection{Specialization VS diversification at a national and international level}

From the times of Adam Smith and David Ricardo, it is recognized that international trade increases specialization, leading to bilateral benefits, despite countries’ differences in technologies and wages. This vision can explain fairly well the inter-industry trade, but it leaves out one of the most important trade modes: the intra-industry trade. The modern approaches aim to solve this by introducing the love for variety \cite{dixit1977,krugman1980}. Indeed diversification of the product basket has been shown to be a good symptom of economic development: developed countries export more intensively and also export a wider basket of goods than their developing counterparts \cite{hummels2005}.
 
The very first analyses of the bipartite representation of the WTW showed an unexpected triangular shape of the biadjacency matrix \cite{Hausmann2007,Hausmann2010}. Indeed, the result was striking, since it showed for the first time the presence of a tendency of countries to diversification: even the most developed countries do not abandon the production of the most basic products, while they enlarge their export basket towards most sophisticated goods. Actually, the picture is less simple than that: although countries tend to diversify their exports, a signal of specialization is still present. Otherwise stated, the observed level of diversification (quantified by the country node degrees) cannot explain a residual tendency of firms to focus on certain classes of products \cite{Straka2017a}.

In order to capture the productive capabilities of countries, several measures were proposed: the very first proposals \cite{Hausmann2007,Hausmann2010} show several flaws and defects \cite{Caldarelli2012}. The Fitness and Complexity algorithm~\cite{Cristelli2013} (\emph{FiCo}, in the following) outperforms other competitors in terms of accuracy of predictions \cite{Cristelli2015a}. The FiCo procedure accounts for the non-linearity of the system via a recursive algorithm: the performance of a country (quantified by a \emph{fitness}) depends both on the ``quality'' of the exported products (described by shades of \emph{complexity}) and on the fitness of their exporters. Indeed, the success of the FiCo algorithm relies upon the structure of the bipartite WTW representation: the algorithm rewards countries according to the variety and complexity of their export baskets.

\begin{figure}[t!]
\includegraphics[width=.6\textwidth]{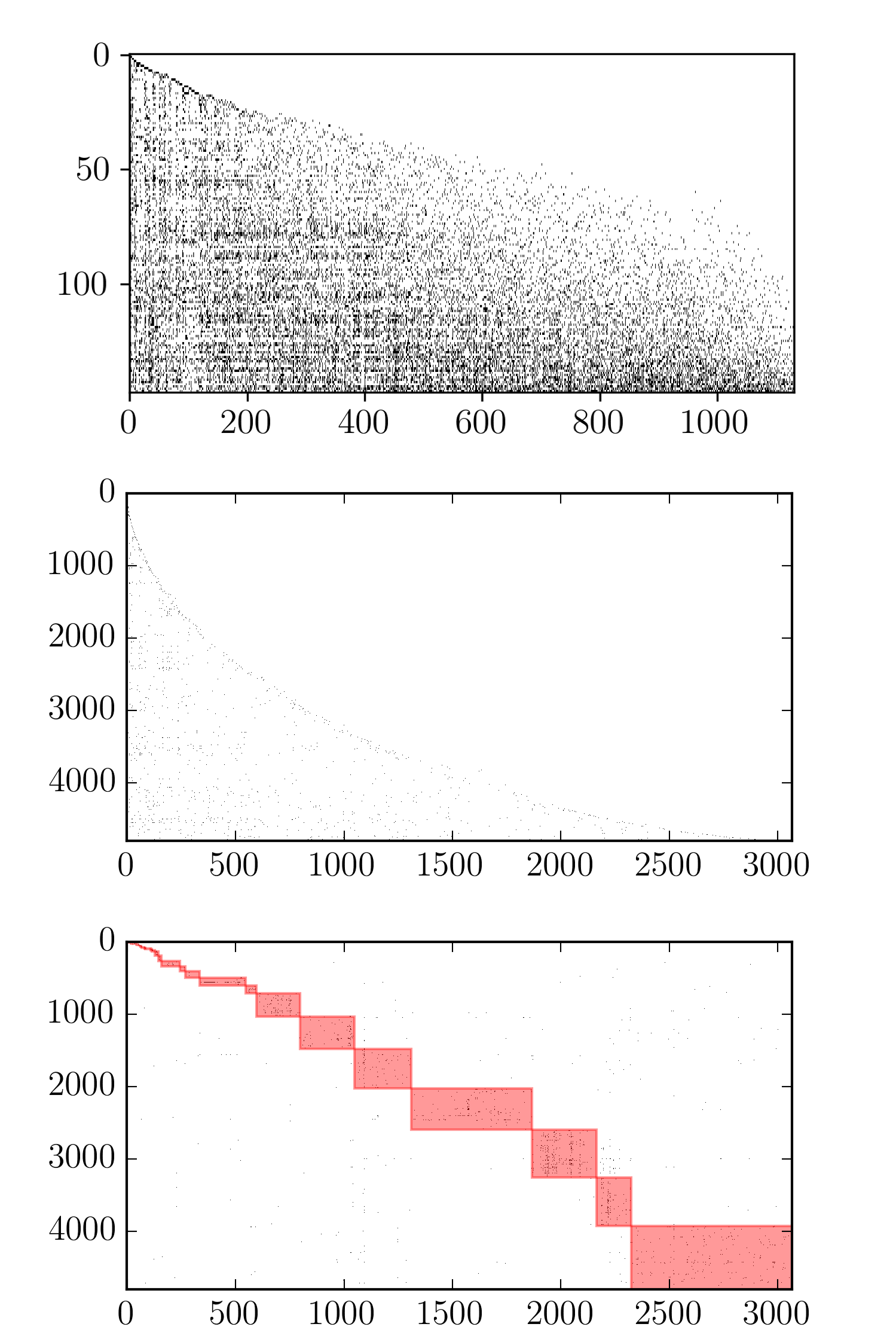}
\caption{Biadjacency matrices of the WTW in the year 2010 (top panel) and of the CPF in the year 2010 (central and bottom panel). The ratio between the x- and y-axes was modify to permit an easier comparison between the shape o the matrices. Columns represent products and rows countries (WTW) or firms (CFP). In the central panel, rows and columns of the biadjacency matrix are ordered according to the FiCo ranking, while in the bottom panel the (bipartite) communities found via the Barber algorithm are highlighted \cite{Barber2007}. The FiCo algorithm, thus, hides the block-structure characterizing the national exports of Colombia.}
\label{figbiadj}
\end{figure}

The FiCo algorithm is also able to highlight the ``triangularity'' of the biadjacency matrix describing a given system \cite{Munoz2013}: as fig. \ref{figbiadj} shows, if rows are re-ordered by fitness and columns by complexity, the non-zero entries of the biadjacency matrix appear as ``packed'' together. When considering the WTW this becomes particularly evident, as the top panel of fig. \ref{figbiadj} shows; analogously, when the FiCo algorithm is applied to the Colombian national export dataset, a triangular structure is revealed as well, as shown in the central panel of fig. \ref{figbiadj} \footnote{In this case, as discussed in \cite{Pugliese2016a} the FiCo algorithm does not converge, but the relative rankings are stable. Actually, only the rankings are necessary for reordering the biadjacency matrix.}.

Interestingly, the triangular structure revealed by the fitness- and complexity-induced re-ordering co-exists with a quite different structure. Upon running the Barber community detection algorithm \cite{Barber2007,Fortunato2010}, based on the bipartite extension of Newman's modularity, a block-wise structure, in fact, emerges (see the bottom panel of fig. \ref{figbiadj}): in a sense, thus, the FiCo algorithm covers the specialization signal present in the original dataset. While at the firm level we indeed expect to observe specialization, since it is unlikely that a company may export all possible products (or even a large percentage of them), this is not true for the WTW, to which the application the same algorithm does not lead to detect any block-wise structure.\\

\begin{figure}[t!]
\includegraphics[width=0.45\textwidth]{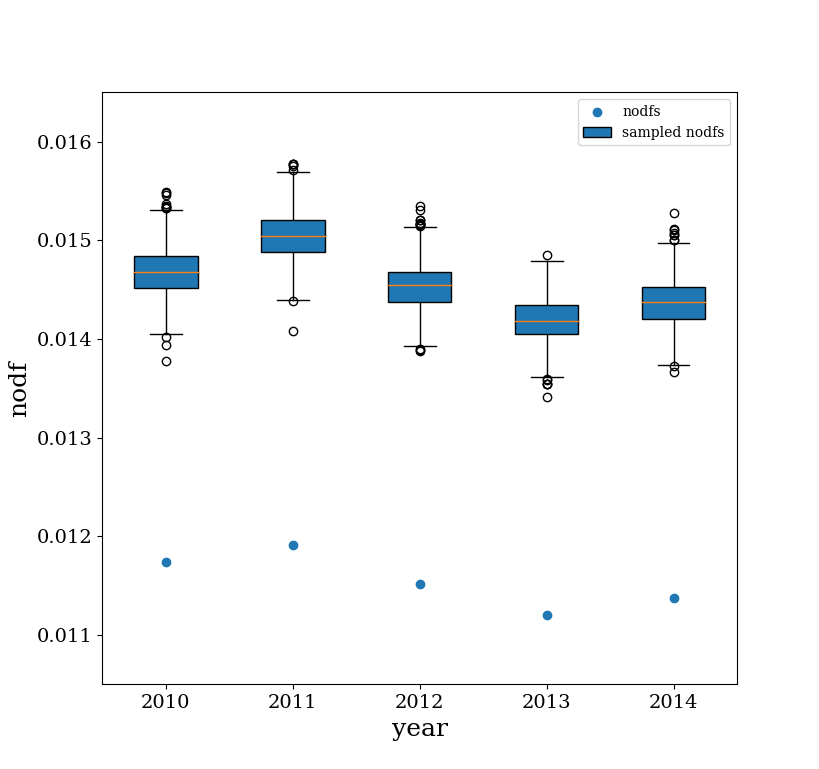}
\caption{Evolution of the empirical nestedness (NODF) values and of the BiCM-induced ensemble distributions of the same quantity, compactly represented by the box-plots (showing the 0.15th, the 25th, the 50th, the 75th and the 99.85th percentiles). The CFP network is characterized by a nestedness whose empirical value is significantly less than expected.}
\label{fignes}
\end{figure}

\paragraph*{Nestedness.} The analysis of nestedness (here we adopted the so-called NODF, i.e. \emph{Nestedness metric based on Overlap and Decreasing Fill} \cite{Almeida-Neto2008a}) allows the picture provided by the FiCo algorithm to be further refined. As fig. \ref{fignes} shows, the z-score of nestedness is steadily negative across our temporal snapshots, i.e. $z_{nestedness}\simeq -11$: in other words, the observed CFP network configurations are significantly less nested than expected, a result that confirms our previous finding concerning the (bipartite) block-structure of the system under analysis. This result further points out that constraining the nodes degrees leads to enforcing some kind of nestedness as well, whose value is (significantly) larger than the observed one.

\subsection{Projecting the Colombian firms-products network}

Let us now move to considering the projections of the CFP network. This kind of analysis complements the results found by running the bipartite community detection shown in fig. \ref{figbiadj}, by making the hidden relationships between nodes belonging to the same layer explicit.

\begin{figure}[hb!]
\includegraphics[width=0.49\textwidth]{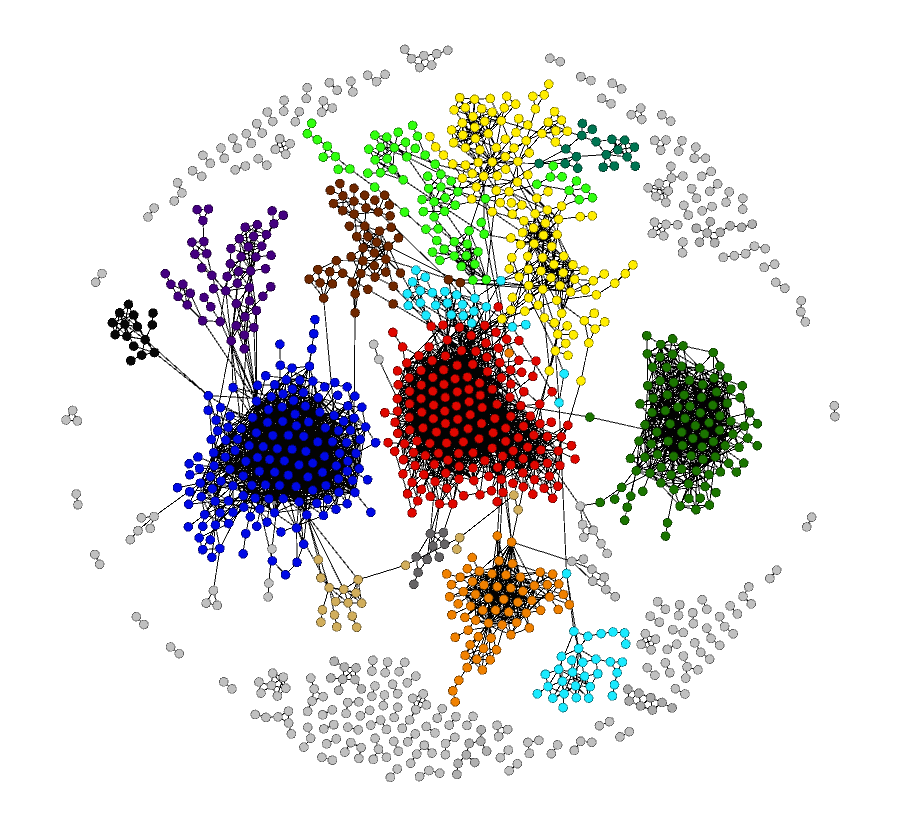}
\includegraphics[width=0.49\textwidth]{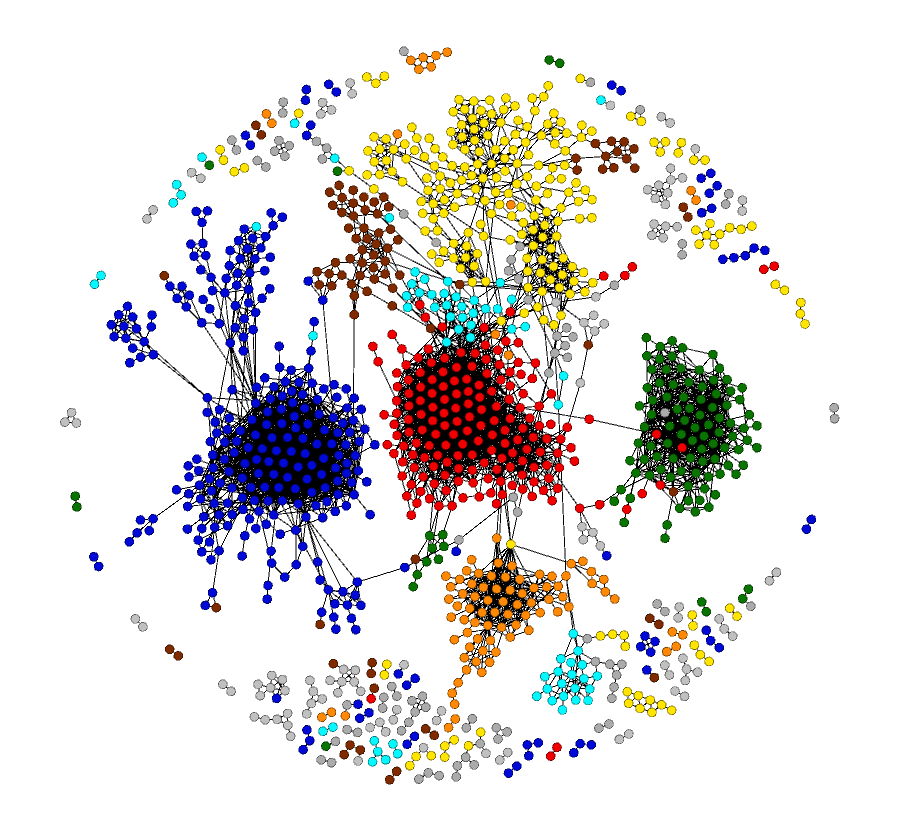}
\includegraphics[width=0.47\textwidth]{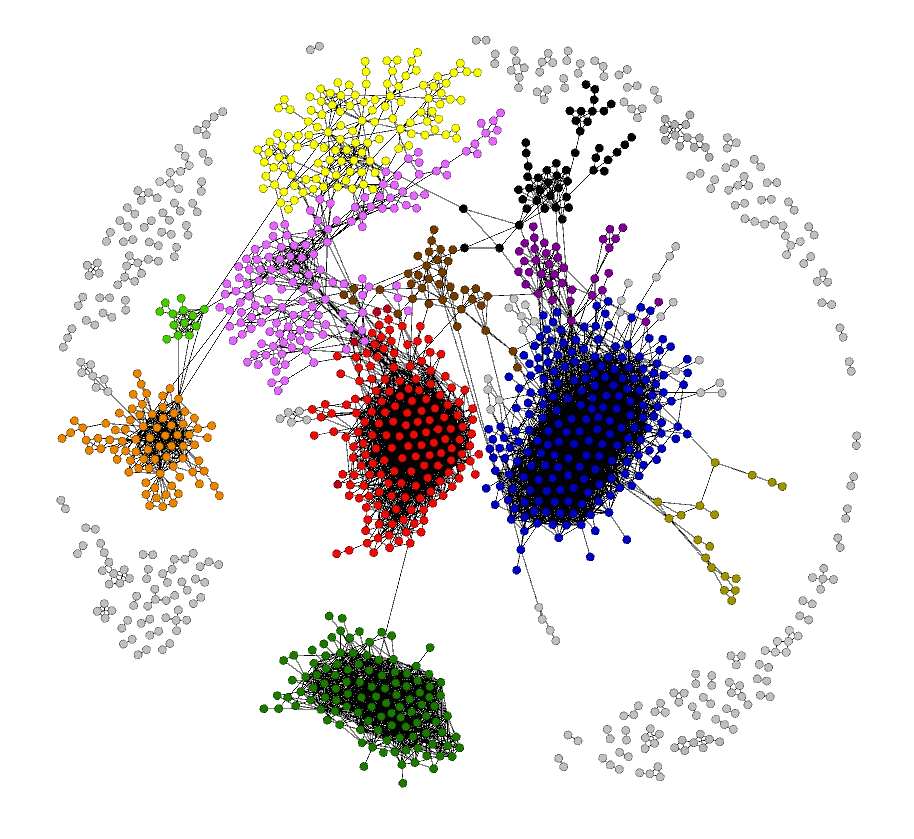}
\includegraphics[width=0.47\textwidth]{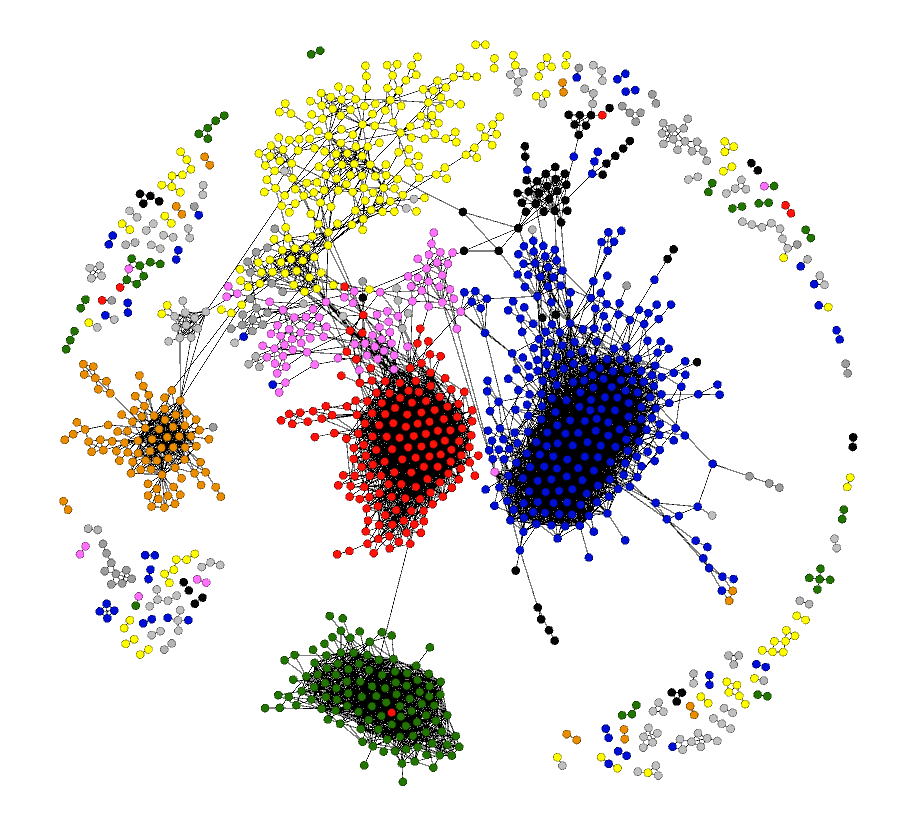}
\caption{Projection of the CFP network on the products layer and detected communities for the years 2010 (top panel) and 2014 (bottom panel), found after the projection of the network (left panel) or found via the Barber algorithm on the original bipartite network, and then projected (right panel). The legend used for the communities on the left is as follows: $\textcolor{red}{\bullet}$ - clothes; $\textcolor{blue}{\bullet}$ - fuels, metals and other industrial products; $\textcolor{OliveGreen}{\bullet}$ - fabrics; $\textcolor{yellow}{\bullet}$ - soaps, body care products and related chemicals; $\textcolor{orange}{\bullet}$ - food; $\textcolor{violet}{\bullet}$ - electronic components; $\textcolor{LimeGreen}{\bullet}$ - chemicals and medicines; $\textcolor{Brown}{\bullet}$ - furniture for the house and ornaments, in wood and plastic; $\textcolor{VioletRed}{\bullet}$ - domestic products, small plastic/metal objects; $\textcolor{SkyBlue}{\bullet}$ - stationery, mixed printed products and kids' toys; $\textcolor{black}{\bullet}$ - small tools for construction companies (chains, hammers, etc.); $\textcolor{Tan}{\bullet}$ - refrigerators and other domestic appliances; $\textcolor{PineGreen}{\bullet}$ - stone, marble and chemicals for construction companies; $\textcolor{Gray}{\bullet}$ - bed linens.}
\label{figprod}
\end{figure}

For what concerns the projection of the CFP network on the layer of products, a persistent structure is observable throughout the years, with the main communities remaining approximately the same (see fig. \ref{figprod}, showing the projections corresponding to the years 2010, 2014 and the related partitions). More in detail, while the total number of connected components is always $\simeq 100$ (some of them are so small that can be neglected for all practical purposes), a number of larger connected components (1-3), characterized by an internal community structure, is observable: additionally, while smaller communities are more homogeneous, the larger ones are more heterogeneous. In any case, the following clusters of products are observed across all temporal snapshots: \emph{clothes}, \emph{industrial supplies}, \emph{bodycare products and related chemicals}, \emph{fabrics and textiles}, \emph{cotton fabrics}, \emph{food}, \emph{electronic devices}, \emph{metal products}, \emph{construction companies supplies}, \emph{domestic appliances}, \emph{leather and footwear}, \emph{stationery}, \emph{wood and glass objects}.

It is also interesting to notice how different communities, characterizing the CFP network projected on the products layer and partitioning the same connected component, are linked. Some examples follow: in 2011 and in 2014, the ``food'' community and the ``bodycare'' community are connected through the product ``organic soaps and essential oils''; in 2013, instead, the ``food'' community is linked to the ``medicines and other chemicals'' community through the product ``vitamins''; in 2014 the ``fabrics'' community is connected to the ``clothes'' community by a single link, joining the ``girls' undershirts'' product and the ``knitted fabrics'' product.

Comparing the communities found in this way on the projected network, our partition is pretty consistent with the one found via the Barber algorithm on the original bipartite network (fig. \ref{figprod}). The Barber communities are usually made up by one or more of the projected communities, plus some of the other nodes not belonging to the giant component (they are indeed connected to their community in the original network). The correlation between the two partitions is measured by the Variation of Information \cite{meila2003}, that in our case is between 0.52 and 0.58 for all five years considered and for both firms and products networks.
Since the Variation of Information is not easy to interpret, it is also possible to measure the fraction of nodes that get included from one of our communities in a Barber one, making a ``best correspondence" between them. We measure the inclusion of our partition $\mathcal{A}$ into the Barber partition $\mathcal{B}$ as
\begin{equation} 
\text{inc}(\mathcal{A}, \mathcal{B}) = \frac{\sum\limits_{A_i \in \mathcal{A}}\max\limits_{B_j \in \mathcal{B}} \vert A_i\cap B_j \vert}{\sum\limits_{A_i \in \mathcal{B}}\vert A_i \vert}
\end{equation}
In our case, this quantity is between 0.85 and 0.92 for all of the years, which means that about 90\% of the nodes are in a community that is consistent  in the two partitions.

\begin{figure}[hb!]
\includegraphics[width=0.49\textwidth]{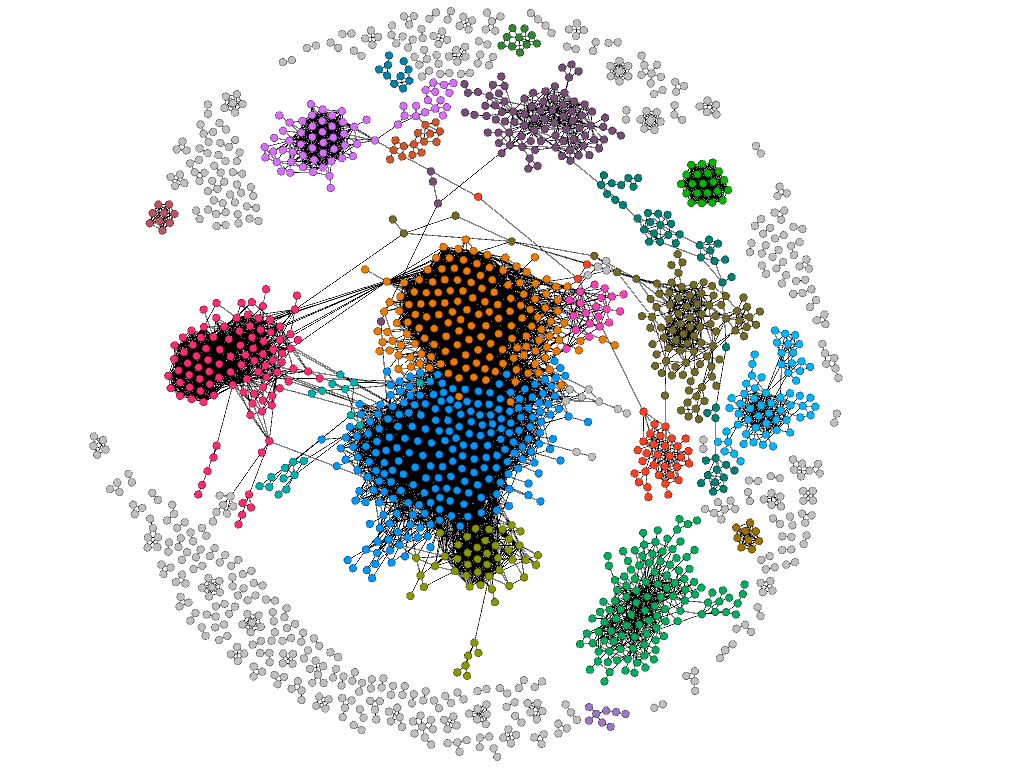}
\includegraphics[width=0.49\textwidth]{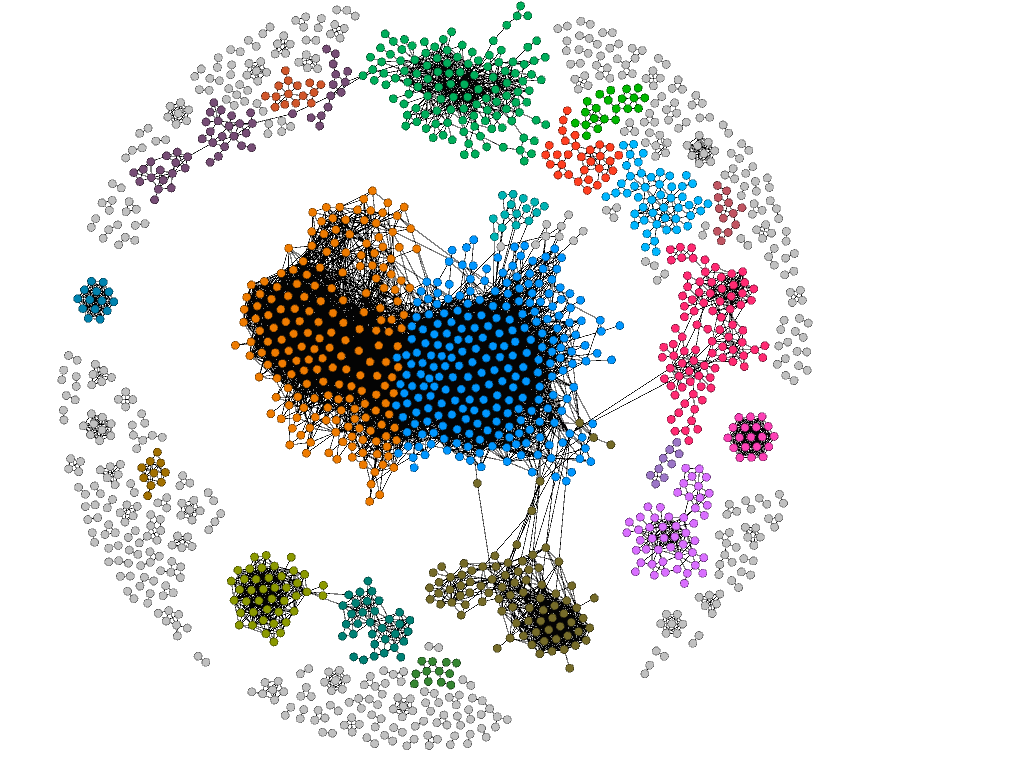}
\caption{Projection of the CFP network on the firms layer and detected communities for the years 2010 (left panel) and 2014 (right panel).}
\label{figfirms}
\end{figure}

The projected CFP network on the firms layer is characterized by a persistent structure throughout our temporal interval as well: this is however denser and composed by larger, isolated components than the projection on the products layer (see fig. \ref{figfirms}).

\paragraph*{Comparison with the WTW.} Let us conclude this section with some remarks: in \cite{Saracco2016b,Straka2017a} the validated projection of the WTW was presented. While the BiCM-induced validation on the layer of countries outputs a clear structure in which countries sharing similar productive capabilities gather in communities, it is necessary to relax the constraints defining the entropy maximization to have a similar projection on the products layer \cite{Saracco2016b,Straka2017a}. From an information-theoretic point of view, the imposed constraints seem to be enough to explain the actual co-occurrences between products: this may be due to the large asymmetry between the cardinality of the countries- and the products-layer, letting the heterogeneity of countries degrees encode all relevant information \cite{Saracco2016b}. Employing a less complex null model, in fact, leads to a projection with a rich structure whose communities of products can be related to the industrialization level of the related exporters: in other words, communities are not defined by homogeneous products but by those that can be efficiently exported by countries with strong industrial capabilities (e.g. metal products, tramway locomotives, tires, and turbines belong to the same cluster).

The case of the national exports of Colombia is essentially different in two main respects: first, we used the BiCM as a benchmark for the projection on both layers; second, the product categories are more clearly defined. This behavior is partly due to the block-wise structure of the system, as already noticed in the previous sections. Finally, let us stress that an intrinsic difference between the bipartite blocks detected by the Barber algorithm and the validated communities characterizing the projections exists. Barber's bipartite modularity compares the local link density with its expected value, thus considering as contributions to the network community structure even small, although positive, fluctuations; the detection of communities on the projections is, instead, enhanced by the preliminary validation implemented via the algorithm introduced in \cite{Saracco2016b}.

\section{Discussion}

\begin{figure}[ht!]
\includegraphics[width=0.45\textwidth]{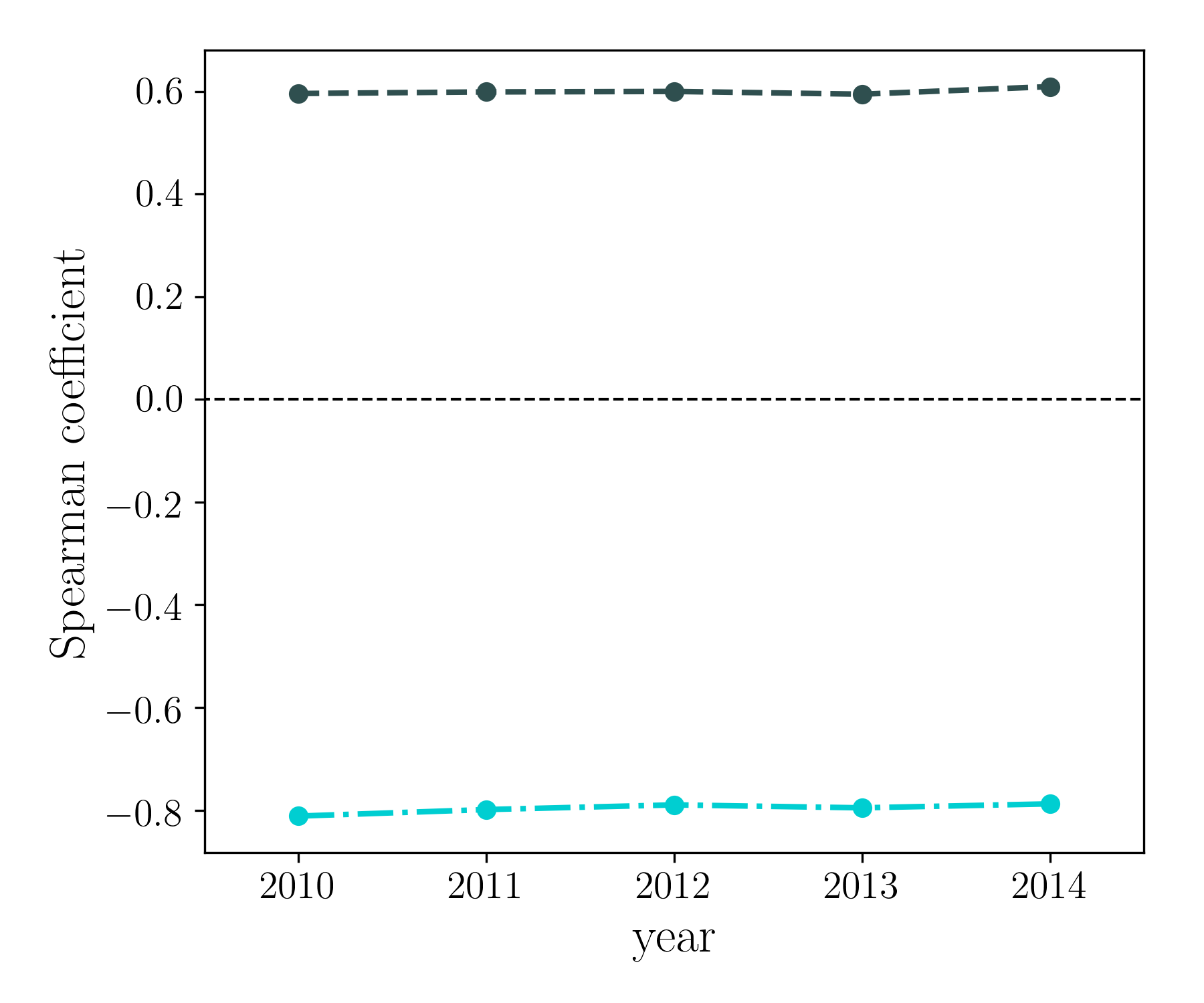}
\includegraphics[width=0.45\textwidth]{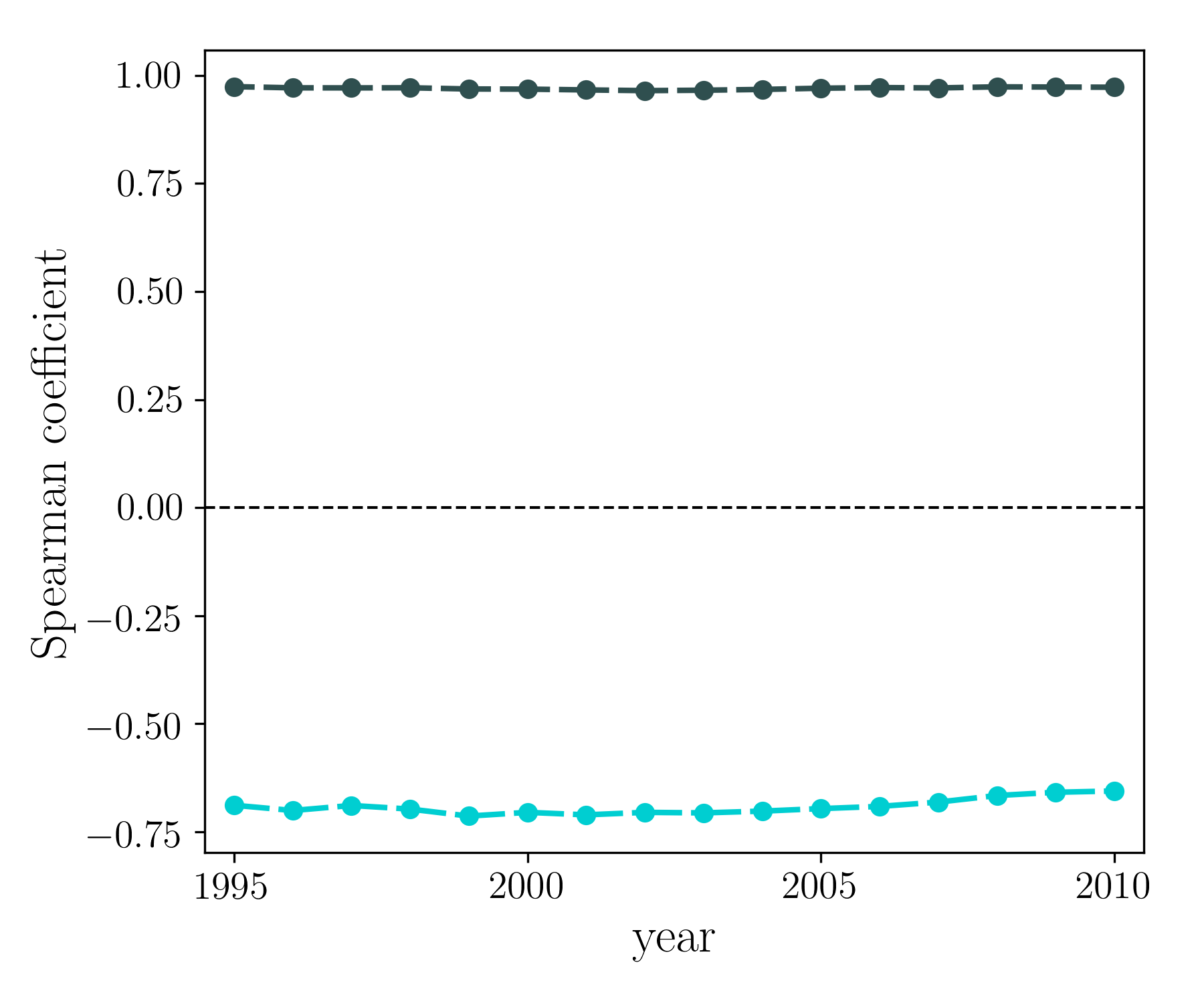}
\caption{Evolution of the Spearman correlation coefficient between the degree of Colombian firms and their fitness values (dark green, dashed, left panel) and between products degree and complexity values (light green, dot-dashed, left panel). As a comparison, the Spearman correlation coefficient between the degree of countries and their fitness values (dark green, dashed, right panel) and between products degree and complexity values (light green, dot-dashed, right panel) is shown.}
\label{figfico}
\end{figure}

In this article we performed a pioneering analysis of the international trade patterns using country firm-level information and employing a complex network methodology. More precisely, we studied the bipartite Colombian firms-products network, using an approach based on the maximization of the constrained Shannon entropy with the available information about the system at hand. This allowed us to detect interesting patterns of economic activities characterizing Colombian firms and products. With the aim of better understanding our results, we kept as a benchmark what we know about the World Trade Web, whose bipartite network (countries-products) has been widely studied in recent years \cite{Mastrandrea2014,Saracco2016b,Straka2017a}.

Both systems have remarkable different organizational structures. The matrix associated with the CFP network is comparatively much sparser than the matrix associated with the WTW network. In fact, while in the WTW network the industrialized countries have the capabilities to export a large number of diverse products, this would be impossible even for the largest companies in Colombia. We showed that the matrix of the CFP network can be re-organized in a block-wise fashion, suggesting the presence of firms specialized in the production of product sets. This is in sharp contrast with the WTW network, whose matrix representation has a genuinely triangular structure.

In this way, we obtained projections in which it was possible to distinguish that both firms and products were organized in communities. We also explicitly notice that the product communities observed in the CFP resemble the WTW communities (although the latter are composed by less homogeneous products than the former ones), being the main similarity between these networks. Although it may be obvious for the economic thinking, it turns out to be interesting verifying that non observable capabilities to produce different products can be recognized at different levels of aggregation, i.e. at both the firm and country levels.

The evidence that diversifying is much more difficult for firms than for countries, in turn leading to little competition at the Colombian, national scale, also affects the performance of the FiCo algorithm in providing information about the ``quality'' of firms and products. Indeed, the FiCo algorithm takes advantage of the triangular shape of the biadjacency matrix. Indeed, it is able to go beyond the degree sequence and highlight those countries that are able to export items that only countries with similar factor endowments and technological capabilities are able to export. In the case of the CFP, there is another crucial point, which is the strong bipartite community structure. Even going beyond the degree sequence, still the community structure is not taken into account. Because of this, the predictions of the FiCo algorithm are going to be weaker on the CFP, with respect to those on the WTW. This is particularly evident when comparing the correlation between the firms' degree and fitness with that of countries degree and fitness (and analogously for what concerns the products - see fig. \ref{figfico}).

Generally speaking, the different behavior of economic systems at different scales is reminiscent of the behavior of ecosystems, with more massive species being characterized by a larger metabolic activity. From this point of view, countries behave like massive species, capable of diversifying their production: firms, on the other hand, are characterized by a much more limited activity, focusing on sectors of products. Indeed, this reflects into the different topological structure of the considered systems, as proven by nestedness, whose observed value is compatible with the BiCM prediction in the case of the WTW \cite{Saracco2015}) but is not for the Colombian national export dataset.\\
The present paper is intended to be the first step of a more extended analysis of national export trade networks. Indeed, due to the recent interest in local nestedness \cite{Sole-Ribalta2018}, it is intriguing the possibility of the analysis in greater details of the structure of the blocks found in fig.\ref{figbiadj}. We leave these studies for further research, due to the implications on the understanding of the whole system.
\section*{Acknowledgements}

This work of F.S. and T. S. was supported by the EU projects CoeGSS (Grant No. 676547), Openmaker (Grant No. 687941), SoBigData (Grant No. 654024). 

\section*{Author Contributions}

F.S., T.S. and M.D. designed the analysis. M.D. provided the data set. M.B. performed the analysis. F.S., T.S. and M.D. wrote the manuscript. All authors reviewed and approved the manuscript.

\section*{Additional Information}

The authors declare no competing financial interests.

\bibliography{Bibliography}

\begin{thebibliography}{48}%
\makeatletter
\providecommand \@ifxundefined [1]{%
 \@ifx{#1\undefined}
}%
\providecommand \@ifnum [1]{%
 \ifnum #1\expandafter \@firstoftwo
 \else \expandafter \@secondoftwo
 \fi
}%
\providecommand \@ifx [1]{%
 \ifx #1\expandafter \@firstoftwo
 \else \expandafter \@secondoftwo
 \fi
}%
\providecommand \natexlab [1]{#1}%
\providecommand \enquote  [1]{``#1''}%
\providecommand \bibnamefont  [1]{#1}%
\providecommand \bibfnamefont [1]{#1}%
\providecommand \citenamefont [1]{#1}%
\providecommand \href@noop [0]{\@secondoftwo}%
\providecommand \href [0]{\begingroup \@sanitize@url \@href}%
\providecommand \@href[1]{\@@startlink{#1}\@@href}%
\providecommand \@@href[1]{\endgroup#1\@@endlink}%
\providecommand \@sanitize@url [0]{\catcode `\\12\catcode `\$12\catcode
  `\&12\catcode `\#12\catcode `\^12\catcode `\_12\catcode `\%12\relax}%
\providecommand \@@startlink[1]{}%
\providecommand \@@endlink[0]{}%
\providecommand \url  [0]{\begingroup\@sanitize@url \@url }%
\providecommand \@url [1]{\endgroup\@href {#1}{\urlprefix }}%
\providecommand \urlprefix  [0]{URL }%
\providecommand \Eprint [0]{\href }%
\providecommand \doibase [0]{http://dx.doi.org/}%
\providecommand \selectlanguage [0]{\@gobble}%
\providecommand \bibinfo  [0]{\@secondoftwo}%
\providecommand \bibfield  [0]{\@secondoftwo}%
\providecommand \translation [1]{[#1]}%
\providecommand \BibitemOpen [0]{}%
\providecommand \bibitemStop [0]{}%
\providecommand \bibitemNoStop [0]{.\EOS\space}%
\providecommand \EOS [0]{\spacefactor3000\relax}%
\providecommand \BibitemShut  [1]{\csname bibitem#1\endcsname}%
\let\auto@bib@innerbib\@empty
\bibitem [{\citenamefont {Hidalgo}\ \emph {et~al.}(2007)\citenamefont
  {Hidalgo}, \citenamefont {Klinger}, \citenamefont {Barab{\'a}si},\ and\
  \citenamefont {Hausmann}}]{Hidalgo2007}%
  \BibitemOpen
  \bibfield  {author} {\bibinfo {author} {\bibfnamefont {C.~A.}\ \bibnamefont
  {Hidalgo}}, \bibinfo {author} {\bibfnamefont {B.}~\bibnamefont {Klinger}},
  \bibinfo {author} {\bibfnamefont {A.-L.}\ \bibnamefont {Barab{\'a}si}}, \
  and\ \bibinfo {author} {\bibfnamefont {R.}~\bibnamefont {Hausmann}},\ }\href
  {\doibase 10.1126/science.1144581} {\bibfield  {journal} {\bibinfo  {journal}
  {Science}\ }\textbf {\bibinfo {volume} {317}},\ \bibinfo {pages} {482}
  (\bibinfo {year} {2007})}\BibitemShut {NoStop}%
\bibitem [{\citenamefont {Caldarelli}\ \emph {et~al.}(2012)\citenamefont
  {Caldarelli}, \citenamefont {Cristelli}, \citenamefont {Gabrielli},
  \citenamefont {Pietronero}, \citenamefont {Scala},\ and\ \citenamefont
  {Tacchella}}]{Caldarelli2012}%
  \BibitemOpen
  \bibfield  {author} {\bibinfo {author} {\bibfnamefont {G.}~\bibnamefont
  {Caldarelli}}, \bibinfo {author} {\bibfnamefont {M.}~\bibnamefont
  {Cristelli}}, \bibinfo {author} {\bibfnamefont {A.}~\bibnamefont
  {Gabrielli}}, \bibinfo {author} {\bibfnamefont {L.}~\bibnamefont
  {Pietronero}}, \bibinfo {author} {\bibfnamefont {A.}~\bibnamefont {Scala}}, \
  and\ \bibinfo {author} {\bibfnamefont {A.}~\bibnamefont {Tacchella}},\ }\href
  {\doibase 10.1371/journal.pone.0047278} {\bibfield  {journal} {\bibinfo
  {journal} {PLoS ONE}\ }\textbf {\bibinfo {volume} {7}},\ \bibinfo {pages}
  {e47278} (\bibinfo {year} {2012})}\BibitemShut {NoStop}%
\bibitem [{\citenamefont {Saracco}\ \emph {et~al.}(2017)\citenamefont
  {Saracco}, \citenamefont {Straka}, \citenamefont {{Di Clemente}},
  \citenamefont {Gabrielli}, \citenamefont {Caldarelli},\ and\ \citenamefont
  {Squartini}}]{Saracco2016b}%
  \BibitemOpen
  \bibfield  {author} {\bibinfo {author} {\bibfnamefont {F.}~\bibnamefont
  {Saracco}}, \bibinfo {author} {\bibfnamefont {M.~J.}\ \bibnamefont {Straka}},
  \bibinfo {author} {\bibfnamefont {R.}~\bibnamefont {{Di Clemente}}}, \bibinfo
  {author} {\bibfnamefont {A.}~\bibnamefont {Gabrielli}}, \bibinfo {author}
  {\bibfnamefont {G.}~\bibnamefont {Caldarelli}}, \ and\ \bibinfo {author}
  {\bibfnamefont {T.}~\bibnamefont {Squartini}},\ }\href {\doibase
  10.1088/1367-2630/aa6b38} {\bibfield  {journal} {\bibinfo  {journal} {New J.
  Phys.}\ }\textbf {\bibinfo {volume} {19}} (\bibinfo {year} {2017}),\
  10.1088/1367-2630/aa6b38},\ \Eprint {http://arxiv.org/abs/1607.02481}
  {arXiv:1607.02481} \BibitemShut {NoStop}%
\bibitem [{\citenamefont {Penrose}(1959)}]{penrose1959}%
  \BibitemOpen
  \bibfield  {author} {\bibinfo {author} {\bibfnamefont {E.}~\bibnamefont
  {Penrose}},\ }\href@noop {} {\emph {\bibinfo {title} {The Theory of the
  Growth of the Firm}}}\ (\bibinfo  {publisher} {Oxford University Press},\
  \bibinfo {address} {Oxford},\ \bibinfo {year} {1959})\BibitemShut {NoStop}%
\bibitem [{\citenamefont {Panzar}\ and\ \citenamefont
  {Willig}(1981)}]{panzar_willing_81}%
  \BibitemOpen
  \bibfield  {author} {\bibinfo {author} {\bibfnamefont {J.~C.}\ \bibnamefont
  {Panzar}}\ and\ \bibinfo {author} {\bibfnamefont {R.~D.}\ \bibnamefont
  {Willig}},\ }\href {http://www.jstor.org/stable/1815729} {\bibfield
  {journal} {\bibinfo  {journal} {The American Economic Review}\ }\textbf
  {\bibinfo {volume} {71}},\ \bibinfo {pages} {268} (\bibinfo {year}
  {1981})}\BibitemShut {NoStop}%
\bibitem [{\citenamefont {Teece}(1980)}]{teece1980}%
  \BibitemOpen
  \bibfield  {author} {\bibinfo {author} {\bibfnamefont {D.~J.}\ \bibnamefont
  {Teece}},\ }\href@noop {} {\bibfield  {journal} {\bibinfo  {journal} {Journal
  of Economic Behavior \& Organization}\ }\textbf {\bibinfo {volume} {1}},\
  \bibinfo {pages} {223} (\bibinfo {year} {1980})}\BibitemShut {NoStop}%
\bibitem [{\citenamefont {Teece}(1982)}]{teece1982}%
  \BibitemOpen
  \bibfield  {author} {\bibinfo {author} {\bibfnamefont {D.~J.}\ \bibnamefont
  {Teece}},\ }\href@noop {} {\bibfield  {journal} {\bibinfo  {journal} {Journal
  of Economic Behavior \& Organization}\ }\textbf {\bibinfo {volume} {3}},\
  \bibinfo {pages} {39} (\bibinfo {year} {1982})}\BibitemShut {NoStop}%
\bibitem [{\citenamefont {Teece}\ \emph {et~al.}(1994)\citenamefont {Teece},
  \citenamefont {Rumelt}, \citenamefont {Dosi},\ and\ \citenamefont
  {Winter}}]{teece1994}%
  \BibitemOpen
  \bibfield  {author} {\bibinfo {author} {\bibfnamefont {D.}~\bibnamefont
  {Teece}}, \bibinfo {author} {\bibfnamefont {R.}~\bibnamefont {Rumelt}},
  \bibinfo {author} {\bibfnamefont {G.}~\bibnamefont {Dosi}}, \ and\ \bibinfo
  {author} {\bibfnamefont {S.}~\bibnamefont {Winter}},\ }\href@noop {}
  {\bibfield  {journal} {\bibinfo  {journal} {Journal of Economic Behavior \&
  Organization}\ }\textbf {\bibinfo {volume} {23}},\ \bibinfo {pages} {1}
  (\bibinfo {year} {1994})}\BibitemShut {NoStop}%
\bibitem [{\citenamefont {Hausmann}\ \emph {et~al.}(2007)\citenamefont
  {Hausmann}, \citenamefont {Hwang},\ and\ \citenamefont
  {Rodrik}}]{Hausmann2007}%
  \BibitemOpen
  \bibfield  {author} {\bibinfo {author} {\bibfnamefont {R.}~\bibnamefont
  {Hausmann}}, \bibinfo {author} {\bibfnamefont {J.}~\bibnamefont {Hwang}}, \
  and\ \bibinfo {author} {\bibfnamefont {D.}~\bibnamefont {Rodrik}},\ }\href
  {\doibase 10.1007/s10887-006-9009-4} {\bibfield  {journal} {\bibinfo
  {journal} {Journal of Economic Growth}\ }\textbf {\bibinfo {volume} {12}},\
  \bibinfo {pages} {1} (\bibinfo {year} {2007})}\BibitemShut {NoStop}%
\bibitem [{\citenamefont {Tacchella}\ \emph {et~al.}(2012)\citenamefont
  {Tacchella}, \citenamefont {Cristelli}, \citenamefont {Caldarelli},
  \citenamefont {Gabrielli},\ and\ \citenamefont {Pietronero}}]{Tacchella2012}%
  \BibitemOpen
  \bibfield  {author} {\bibinfo {author} {\bibfnamefont {A.}~\bibnamefont
  {Tacchella}}, \bibinfo {author} {\bibfnamefont {M.}~\bibnamefont
  {Cristelli}}, \bibinfo {author} {\bibfnamefont {G.}~\bibnamefont
  {Caldarelli}}, \bibinfo {author} {\bibfnamefont {A.}~\bibnamefont
  {Gabrielli}}, \ and\ \bibinfo {author} {\bibfnamefont {L.}~\bibnamefont
  {Pietronero}},\ }\href {http://dx.doi.org/10.1038/srep00723} {\bibfield
  {journal} {\bibinfo  {journal} {Scientific Reports}\ }\textbf {\bibinfo
  {volume} {2}},\ \bibinfo {pages} {723} (\bibinfo {year} {2012})}\BibitemShut
  {NoStop}%
\bibitem [{\citenamefont {Cristelli}\ \emph {et~al.}(2013)\citenamefont
  {Cristelli}, \citenamefont {Gabrielli}, \citenamefont {Tacchella},
  \citenamefont {Caldarelli},\ and\ \citenamefont
  {Pietronero}}]{Cristelli2013}%
  \BibitemOpen
  \bibfield  {author} {\bibinfo {author} {\bibfnamefont {M.}~\bibnamefont
  {Cristelli}}, \bibinfo {author} {\bibfnamefont {A.}~\bibnamefont
  {Gabrielli}}, \bibinfo {author} {\bibfnamefont {A.}~\bibnamefont
  {Tacchella}}, \bibinfo {author} {\bibfnamefont {G.}~\bibnamefont
  {Caldarelli}}, \ and\ \bibinfo {author} {\bibfnamefont {L.}~\bibnamefont
  {Pietronero}},\ }\href@noop {} {\bibfield  {journal} {\bibinfo  {journal}
  {PLoS One}\ }\textbf {\bibinfo {volume} {8}} (\bibinfo {year}
  {2013})}\BibitemShut {NoStop}%
\bibitem [{Note1()}]{Note1}%
  \BibitemOpen
  \bibinfo {note} {The latter is the case in which comparative advantages of
  countries induce specialization in a few products according to their factor
  and technological endowments.}\BibitemShut {Stop}%
\bibitem [{Note2()}]{Note2}%
  \BibitemOpen
  \bibinfo {note} {\protect \href
  {https://unstats.un.org/unsd/trade/classifications/correspondence-tables.asp}{https://unstats.un.org/unsd/trade/classifications/correspondence-tables.asp}}\BibitemShut
  {NoStop}%
\bibitem [{\citenamefont {G.~Gaulier}(2013)}]{BACI2013}%
  \BibitemOpen
  \bibfield  {author} {\bibinfo {author} {\bibfnamefont {S.}~\bibnamefont
  {G.~Gaulier}},\ }\href {http://www.cepii.fr/anglaisgraph/workpap/pdf/2010/
  wp2010-23.pdf} {\bibfield  {journal} {\bibinfo  {journal} {BACI:
  International Trade Database at the Product Level}\ } (\bibinfo {year} {last
  access: 5 July 2013})}\BibitemShut {NoStop}%
\bibitem [{\citenamefont {Balassa}(1977)}]{Balassa1977}%
  \BibitemOpen
  \bibfield  {author} {\bibinfo {author} {\bibfnamefont {B.}~\bibnamefont
  {Balassa}},\ }\href@noop {} {\bibfield  {journal} {\bibinfo  {journal} {The
  Manchester School}\ }\textbf {\bibinfo {volume} {45}},\ \bibinfo {pages}
  {327} (\bibinfo {year} {1977})}\BibitemShut {NoStop}%
\bibitem [{\citenamefont {Bottazzi}\ and\ \citenamefont
  {Pirino}(2010)}]{Bottazzi2010}%
  \BibitemOpen
  \bibfield  {author} {\bibinfo {author} {\bibfnamefont {G.}~\bibnamefont
  {Bottazzi}}\ and\ \bibinfo {author} {\bibfnamefont {D.}~\bibnamefont
  {Pirino}},\ }\href {\doibase 10.2139/ssrn.1831479} {\enquote {\bibinfo
  {title} {{Measuring Industry Relatedness and Corporate Coherence}},}\ }
  (\bibinfo {year} {2010})\BibitemShut {NoStop}%
\bibitem [{\citenamefont {Park}\ and\ \citenamefont {Newman}(2004)}]{Park2004}%
  \BibitemOpen
  \bibfield  {author} {\bibinfo {author} {\bibfnamefont {J.}~\bibnamefont
  {Park}}\ and\ \bibinfo {author} {\bibfnamefont {M.~E.~J.}\ \bibnamefont
  {Newman}},\ }\href {\doibase 10.1103/PhysRevE.70.066117} {\bibfield
  {journal} {\bibinfo  {journal} {Physical Review E}\ }\textbf {\bibinfo
  {volume} {70}},\ \bibinfo {pages} {066117} (\bibinfo {year}
  {2004})}\BibitemShut {NoStop}%
\bibitem [{\citenamefont {Garlaschelli}\ and\ \citenamefont
  {Loffredo}(2008)}]{Garlaschelli2008}%
  \BibitemOpen
  \bibfield  {author} {\bibinfo {author} {\bibfnamefont {D.}~\bibnamefont
  {Garlaschelli}}\ and\ \bibinfo {author} {\bibfnamefont {M.~I.}\ \bibnamefont
  {Loffredo}},\ }\href {\doibase 10.1103/PhysRevE.78.015101} {\bibfield
  {journal} {\bibinfo  {journal} {Physical Review E}\ }\textbf {\bibinfo
  {volume} {78}},\ \bibinfo {pages} {015101} (\bibinfo {year}
  {2008})}\BibitemShut {NoStop}%
\bibitem [{\citenamefont {{Squartini}}\ and\ \citenamefont
  {{Garlaschelli}}(2011)}]{Squartini2011}%
  \BibitemOpen
  \bibfield  {author} {\bibinfo {author} {\bibfnamefont {T.}~\bibnamefont
  {{Squartini}}}\ and\ \bibinfo {author} {\bibfnamefont {D.}~\bibnamefont
  {{Garlaschelli}}},\ }\href {\doibase 10.1088/1367-2630/13/8/083001}
  {\bibfield  {journal} {\bibinfo  {journal} {New Journal of Physics}\ }\textbf
  {\bibinfo {volume} {13}},\ \bibinfo {pages} {083001} (\bibinfo {year}
  {2011})}\BibitemShut {NoStop}%
\bibitem [{\citenamefont {Fronczak}(2014)}]{Fronczak2012}%
  \BibitemOpen
  \bibfield  {author} {\bibinfo {author} {\bibfnamefont {A.}~\bibnamefont
  {Fronczak}},\ }\href {\doibase 10.1007/978-1-4614-6170-8} {\bibfield
  {journal} {\bibinfo  {journal} {Encyclopedia of Social Network Analysis and
  Mining}\ ,\ \bibinfo {pages} {500}} (\bibinfo {year} {2014})}\BibitemShut
  {NoStop}%
\bibitem [{\citenamefont {Saracco}\ \emph {et~al.}(2015)\citenamefont
  {Saracco}, \citenamefont {Di~Clemente}, \citenamefont {Gabrielli},\ and\
  \citenamefont {Squartini}}]{Saracco2015}%
  \BibitemOpen
  \bibfield  {author} {\bibinfo {author} {\bibfnamefont {F.}~\bibnamefont
  {Saracco}}, \bibinfo {author} {\bibfnamefont {R.}~\bibnamefont
  {Di~Clemente}}, \bibinfo {author} {\bibfnamefont {A.}~\bibnamefont
  {Gabrielli}}, \ and\ \bibinfo {author} {\bibfnamefont {T.}~\bibnamefont
  {Squartini}},\ }\href {\doibase 10.1038/srep10595} {\bibfield  {journal}
  {\bibinfo  {journal} {Scientific Reports}\ }\textbf {\bibinfo {volume} {5}},\
  \bibinfo {pages} {10595} (\bibinfo {year} {2015})}\BibitemShut {NoStop}%
\bibitem [{\citenamefont {Hong}(2013)}]{Hong2013}%
  \BibitemOpen
  \bibfield  {author} {\bibinfo {author} {\bibfnamefont {Y.}~\bibnamefont
  {Hong}},\ }\href {\doibase 10.1016/j.csda.2012.10.006} {\bibfield  {journal}
  {\bibinfo  {journal} {Computational Statistics And Data Analysis}\ }\textbf
  {\bibinfo {volume} {59}},\ \bibinfo {pages} {41} (\bibinfo {year}
  {2013})}\BibitemShut {NoStop}%
\bibitem [{\citenamefont {Deheuvels}\ \emph {et~al.}(1989)\citenamefont
  {Deheuvels}, \citenamefont {Puri},\ and\ \citenamefont
  {Ralescu}}]{Deheuvels1989}%
  \BibitemOpen
  \bibfield  {author} {\bibinfo {author} {\bibfnamefont {P.}~\bibnamefont
  {Deheuvels}}, \bibinfo {author} {\bibfnamefont {M.~L.}\ \bibnamefont {Puri}},
  \ and\ \bibinfo {author} {\bibfnamefont {S.~S.}\ \bibnamefont {Ralescu}},\
  }\href {\doibase 10.1016/0047-259X(89)90111-5} {\bibfield  {journal}
  {\bibinfo  {journal} {Journal of Multivariate Analysis}\ }\textbf {\bibinfo
  {volume} {28}},\ \bibinfo {pages} {282} (\bibinfo {year} {1989})}\BibitemShut
  {NoStop}%
\bibitem [{\citenamefont {Volkova}(1996)}]{Volkova1996}%
  \BibitemOpen
  \bibfield  {author} {\bibinfo {author} {\bibfnamefont {A.~Y.}\ \bibnamefont
  {Volkova}},\ }\href {\doibase 10.1137/1140093} {\bibfield  {journal}
  {\bibinfo  {journal} {Theory of Probability And Its Applications}\ }\textbf
  {\bibinfo {volume} {40}},\ \bibinfo {pages} {791} (\bibinfo {year}
  {1996})}\BibitemShut {NoStop}%
\bibitem [{\citenamefont {Saracco}\ \emph {et~al.}(2016)\citenamefont
  {Saracco}, \citenamefont {Di~Clemente}, \citenamefont {Gabrielli},\ and\
  \citenamefont {Squartini}}]{Saracco2016}%
  \BibitemOpen
  \bibfield  {author} {\bibinfo {author} {\bibfnamefont {F.}~\bibnamefont
  {Saracco}}, \bibinfo {author} {\bibfnamefont {R.}~\bibnamefont
  {Di~Clemente}}, \bibinfo {author} {\bibfnamefont {A.}~\bibnamefont
  {Gabrielli}}, \ and\ \bibinfo {author} {\bibfnamefont {T.}~\bibnamefont
  {Squartini}},\ }\href {http://dx.doi.org/10.1038/srep30286} {\bibfield
  {journal} {\bibinfo  {journal} {Scientific Reports}\ }\textbf {\bibinfo
  {volume} {6}},\ \bibinfo {pages} {30286} (\bibinfo {year}
  {2016})}\BibitemShut {NoStop}%
\bibitem [{\citenamefont {Benjamini}\ and\ \citenamefont
  {Hochberg}(1995)}]{Benjamini1995}%
  \BibitemOpen
  \bibfield  {author} {\bibinfo {author} {\bibfnamefont {Y.}~\bibnamefont
  {Benjamini}}\ and\ \bibinfo {author} {\bibfnamefont {Y.}~\bibnamefont
  {Hochberg}},\ }\href@noop {} {\bibfield  {journal} {\bibinfo  {journal}
  {Journal of the Royal Statistical Society B}\ }\textbf {\bibinfo {volume}
  {57}},\ \bibinfo {pages} {289} (\bibinfo {year} {1995})}\BibitemShut
  {NoStop}%
\bibitem [{Note3()}]{Note3}%
  \BibitemOpen
  \bibinfo {note} {Our hypotheses, for example, are not independent, since each
  observed link affects the similarity of several pairs of nodes}\BibitemShut
  {NoStop}%
\bibitem [{\citenamefont {Blondel}\ \emph {et~al.}(2008)\citenamefont
  {Blondel}, \citenamefont {Guillaume}, \citenamefont {Lambiotte},\ and\
  \citenamefont {Lefebvre}}]{Blondel2008}%
  \BibitemOpen
  \bibfield  {author} {\bibinfo {author} {\bibfnamefont {V.~D.}\ \bibnamefont
  {Blondel}}, \bibinfo {author} {\bibfnamefont {J.-L.}\ \bibnamefont
  {Guillaume}}, \bibinfo {author} {\bibfnamefont {R.}~\bibnamefont
  {Lambiotte}}, \ and\ \bibinfo {author} {\bibfnamefont {E.}~\bibnamefont
  {Lefebvre}},\ }\href {\doibase 10.1088/1742-5468/2008/10/P10008} {\bibfield
  {journal} {\bibinfo  {journal} {Journal of Statistical Mechanics: Theory and
  Experiment}\ }\textbf {\bibinfo {volume} {2008}},\ \bibinfo {pages} {P10008}
  (\bibinfo {year} {2008})}\BibitemShut {NoStop}%
\bibitem [{\citenamefont {{Fortunato}}(2010)}]{Fortunato2010}%
  \BibitemOpen
  \bibfield  {author} {\bibinfo {author} {\bibfnamefont {S.}~\bibnamefont
  {{Fortunato}}},\ }\href {\doibase 10.1016/j.physrep.2009.11.002} {\bibfield
  {journal} {\bibinfo  {journal} {Physics Reports}\ }\textbf {\bibinfo {volume}
  {486}},\ \bibinfo {pages} {75} (\bibinfo {year} {2010})}\BibitemShut
  {NoStop}%
\bibitem [{\citenamefont {Fagiolo}\ \emph {et~al.}(2009)\citenamefont
  {Fagiolo}, \citenamefont {Schiavo},\ and\ \citenamefont
  {Reyes}}]{fagiolo2009pre}%
  \BibitemOpen
  \bibfield  {author} {\bibinfo {author} {\bibfnamefont {G.}~\bibnamefont
  {Fagiolo}}, \bibinfo {author} {\bibfnamefont {S.}~\bibnamefont {Schiavo}}, \
  and\ \bibinfo {author} {\bibfnamefont {J.}~\bibnamefont {Reyes}},\
  }\href@noop {} {\bibfield  {journal} {\bibinfo  {journal} {Physical Review
  E}\ }\textbf {\bibinfo {volume} {79}},\ \bibinfo {pages} {036115} (\bibinfo
  {year} {2009})}\BibitemShut {NoStop}%
\bibitem [{\citenamefont {Bee}\ \emph {et~al.}(2017)\citenamefont {Bee},
  \citenamefont {Riccaboni},\ and\ \citenamefont {Schiavo}}]{Bee_etal_2017}%
  \BibitemOpen
  \bibfield  {author} {\bibinfo {author} {\bibfnamefont {M.}~\bibnamefont
  {Bee}}, \bibinfo {author} {\bibfnamefont {M.}~\bibnamefont {Riccaboni}}, \
  and\ \bibinfo {author} {\bibfnamefont {S.}~\bibnamefont {Schiavo}},\ }\href
  {\doibase https://doi.org/10.1016/j.physa.2017.04.012} {\bibfield  {journal}
  {\bibinfo  {journal} {Physica A: Statistical Mechanics and its Applications}\
  }\textbf {\bibinfo {volume} {481}},\ \bibinfo {pages} {265 } (\bibinfo {year}
  {2017})}\BibitemShut {NoStop}%
\bibitem [{\citenamefont {{Campi}}\ \emph {et~al.}(2018)\citenamefont
  {{Campi}}, \citenamefont {{Due{\~n}as}}, \citenamefont {{Li}},\ and\
  \citenamefont {{Wu}}}]{Campi_etal_2017}%
  \BibitemOpen
  \bibfield  {author} {\bibinfo {author} {\bibfnamefont {M.}~\bibnamefont
  {{Campi}}}, \bibinfo {author} {\bibfnamefont {M.}~\bibnamefont
  {{Due{\~n}as}}}, \bibinfo {author} {\bibfnamefont {L.}~\bibnamefont {{Li}}},
  \ and\ \bibinfo {author} {\bibfnamefont {H.}~\bibnamefont {{Wu}}},\
  }\href@noop {} {\bibfield  {journal} {\bibinfo  {journal} {ArXiv e-prints}\ }
  (\bibinfo {year} {2018})},\ \Eprint {http://arxiv.org/abs/1801.02681}
  {arXiv:1801.02681} \BibitemShut {NoStop}%
\bibitem [{\citenamefont {Cimini}\ \emph {et~al.}(2015)\citenamefont {Cimini},
  \citenamefont {Squartini}, \citenamefont {Garlaschelli},\ and\ \citenamefont
  {Gabrielli}}]{Cimini2015}%
  \BibitemOpen
  \bibfield  {author} {\bibinfo {author} {\bibfnamefont {G.}~\bibnamefont
  {Cimini}}, \bibinfo {author} {\bibfnamefont {T.}~\bibnamefont {Squartini}},
  \bibinfo {author} {\bibfnamefont {D.}~\bibnamefont {Garlaschelli}}, \ and\
  \bibinfo {author} {\bibfnamefont {A.}~\bibnamefont {Gabrielli}},\ }\href
  {\doibase 10.1038/srep15758} {\bibfield  {journal} {\bibinfo  {journal}
  {Scientific Reports}\ }\textbf {\bibinfo {volume} {5}},\ \bibinfo {pages}
  {15758} (\bibinfo {year} {2015})}\BibitemShut {NoStop}%
\bibitem [{\citenamefont {Squartini}\ \emph {et~al.}(2017)\citenamefont
  {Squartini}, \citenamefont {Almog}, \citenamefont {Caldarelli}, \citenamefont
  {van Lelyveld}, \citenamefont {Garlaschelli},\ and\ \citenamefont
  {Cimini}}]{Squartini2017}%
  \BibitemOpen
  \bibfield  {author} {\bibinfo {author} {\bibfnamefont {T.}~\bibnamefont
  {Squartini}}, \bibinfo {author} {\bibfnamefont {A.}~\bibnamefont {Almog}},
  \bibinfo {author} {\bibfnamefont {G.}~\bibnamefont {Caldarelli}}, \bibinfo
  {author} {\bibfnamefont {I.}~\bibnamefont {van Lelyveld}}, \bibinfo {author}
  {\bibfnamefont {D.}~\bibnamefont {Garlaschelli}}, \ and\ \bibinfo {author}
  {\bibfnamefont {G.}~\bibnamefont {Cimini}},\ }\href {\doibase
  10.1103/PhysRevE.96.032315} {\bibfield  {journal} {\bibinfo  {journal}
  {Physical Review E}\ }\textbf {\bibinfo {volume} {96}},\ \bibinfo {pages}
  {032315} (\bibinfo {year} {2017})}\BibitemShut {NoStop}%
\bibitem [{\citenamefont {Dixit}\ and\ \citenamefont
  {Stiglitz}(1977)}]{dixit1977}%
  \BibitemOpen
  \bibfield  {author} {\bibinfo {author} {\bibfnamefont {A.~K.}\ \bibnamefont
  {Dixit}}\ and\ \bibinfo {author} {\bibfnamefont {J.~E.}\ \bibnamefont
  {Stiglitz}},\ }\href@noop {} {\bibfield  {journal} {\bibinfo  {journal} {The
  American Economic Review}\ }\textbf {\bibinfo {volume} {67}},\ \bibinfo
  {pages} {297} (\bibinfo {year} {1977})}\BibitemShut {NoStop}%
\bibitem [{\citenamefont {Krugman}(1980)}]{krugman1980}%
  \BibitemOpen
  \bibfield  {author} {\bibinfo {author} {\bibfnamefont {P.}~\bibnamefont
  {Krugman}},\ }\href@noop {} {\bibfield  {journal} {\bibinfo  {journal} {The
  American Economic Review}\ }\textbf {\bibinfo {volume} {70}},\ \bibinfo
  {pages} {950} (\bibinfo {year} {1980})}\BibitemShut {NoStop}%
\bibitem [{\citenamefont {Hummels}\ and\ \citenamefont
  {Klenow}(2005)}]{hummels2005}%
  \BibitemOpen
  \bibfield  {author} {\bibinfo {author} {\bibfnamefont {D.}~\bibnamefont
  {Hummels}}\ and\ \bibinfo {author} {\bibfnamefont {P.~J.}\ \bibnamefont
  {Klenow}},\ }\href@noop {} {\bibfield  {journal} {\bibinfo  {journal}
  {American Economic Review}\ }\textbf {\bibinfo {volume} {95}},\ \bibinfo
  {pages} {704} (\bibinfo {year} {2005})}\BibitemShut {NoStop}%
\bibitem [{\citenamefont {Hausmann}\ and\ \citenamefont
  {Hidalgo}(2010)}]{Hausmann2010}%
  \BibitemOpen
  \bibfield  {author} {\bibinfo {author} {\bibfnamefont {R.}~\bibnamefont
  {Hausmann}}\ and\ \bibinfo {author} {\bibfnamefont {C.}~\bibnamefont
  {Hidalgo}},\ }\href@noop {} {\bibfield  {journal} {\bibinfo  {journal} {HKS
  Working Paper No. RWP 10-045}\ } (\bibinfo {year} {2010})}\BibitemShut
  {NoStop}%
\bibitem [{\citenamefont {Straka}\ \emph {et~al.}(2017)\citenamefont {Straka},
  \citenamefont {Caldarelli},\ and\ \citenamefont {Saracco}}]{Straka2017a}%
  \BibitemOpen
  \bibfield  {author} {\bibinfo {author} {\bibfnamefont {M.~J.}\ \bibnamefont
  {Straka}}, \bibinfo {author} {\bibfnamefont {G.}~\bibnamefont {Caldarelli}},
  \ and\ \bibinfo {author} {\bibfnamefont {F.}~\bibnamefont {Saracco}},\ }\href
  {\doibase 10.1103/PhysRevE.96.022306} {\bibfield  {journal} {\bibinfo
  {journal} {Phys. Rev. E}\ }\textbf {\bibinfo {volume} {96}} (\bibinfo {year}
  {2017}),\ 10.1103/PhysRevE.96.022306},\ \Eprint
  {http://arxiv.org/abs/arXiv:1703.04090v1} {arXiv:arXiv:1703.04090v1}
  \BibitemShut {NoStop}%
\bibitem [{\citenamefont {Cristelli}\ \emph {et~al.}(2015)\citenamefont
  {Cristelli}, \citenamefont {Tacchella},\ and\ \citenamefont
  {Pietronero}}]{Cristelli2015a}%
  \BibitemOpen
  \bibfield  {author} {\bibinfo {author} {\bibfnamefont {M.}~\bibnamefont
  {Cristelli}}, \bibinfo {author} {\bibfnamefont {A.}~\bibnamefont
  {Tacchella}}, \ and\ \bibinfo {author} {\bibfnamefont {L.}~\bibnamefont
  {Pietronero}},\ }\href {\doibase 10.1371/journal.pone.0117174} {\bibfield
  {journal} {\bibinfo  {journal} {PLoS One}\ }\textbf {\bibinfo {volume} {10}}
  (\bibinfo {year} {2015}),\ 10.1371/journal.pone.0117174}\BibitemShut
  {NoStop}%
\bibitem [{\citenamefont {{Johnson}}\ \emph {et~al.}(2013)\citenamefont
  {{Johnson}}, \citenamefont {{Dominguez-Garcia}},\ and\ \citenamefont
  {{Munoz}}}]{Munoz2013}%
  \BibitemOpen
  \bibfield  {author} {\bibinfo {author} {\bibfnamefont {S.}~\bibnamefont
  {{Johnson}}}, \bibinfo {author} {\bibfnamefont {V.}~\bibnamefont
  {{Dominguez-Garcia}}}, \ and\ \bibinfo {author} {\bibfnamefont {M.~A.}\
  \bibnamefont {{Munoz}}},\ }\href@noop {} {\bibfield  {journal} {\bibinfo
  {journal} {ArXiv e-prints}\ } (\bibinfo {year} {2013})},\ \Eprint
  {http://arxiv.org/abs/1307.4685} {arXiv:1307.4685 [physics.soc-ph]}
  \BibitemShut {NoStop}%
\bibitem [{Note4()}]{Note4}%
  \BibitemOpen
  \bibinfo {note} {In this case, as discussed in \cite {Pugliese2016a} the FiCo
  algorithm does not converge, but the relative rankings are stable. Actually,
  only the rankings are necessary for reordering the biadjacency
  matrix.}\BibitemShut {Stop}%
\bibitem [{\citenamefont {Barber}(2007)}]{Barber2007}%
  \BibitemOpen
  \bibfield  {author} {\bibinfo {author} {\bibfnamefont {M.~J.}\ \bibnamefont
  {Barber}},\ }\href {\doibase 10.1103/PhysRevE.76.066102} {\bibfield
  {journal} {\bibinfo  {journal} {Phys. Rev. E - Stat. Nonlinear, Soft Matter
  Phys.}\ }\textbf {\bibinfo {volume} {76}} (\bibinfo {year} {2007}),\
  10.1103/PhysRevE.76.066102},\ \Eprint {http://arxiv.org/abs/0707.1616}
  {arXiv:0707.1616} \BibitemShut {NoStop}%
\bibitem [{\citenamefont {Almeida-Neto}\ \emph {et~al.}(2008)\citenamefont
  {Almeida-Neto}, \citenamefont {Guimar{\~{a}}es}, \citenamefont {Guimaraes},
  \citenamefont {Loyola},\ and\ \citenamefont {Ulrich}}]{Almeida-Neto2008a}%
  \BibitemOpen
  \bibfield  {author} {\bibinfo {author} {\bibfnamefont {M.}~\bibnamefont
  {Almeida-Neto}}, \bibinfo {author} {\bibfnamefont {P.}~\bibnamefont
  {Guimar{\~{a}}es}}, \bibinfo {author} {\bibfnamefont {J.~P.~R.}\ \bibnamefont
  {Guimaraes}}, \bibinfo {author} {\bibfnamefont {R.~D.}\ \bibnamefont
  {Loyola}}, \ and\ \bibinfo {author} {\bibfnamefont {W.}~\bibnamefont
  {Ulrich}},\ }\href {\doibase 10.1111/j.0030-1299.2008.16644.x} {\bibfield
  {journal} {\bibinfo  {journal} {Oikos}\ }\textbf {\bibinfo {volume} {117}},\
  \bibinfo {pages} {1227} (\bibinfo {year} {2008})}\BibitemShut {NoStop}%
\bibitem [{\citenamefont {Meil{\u{a}}}(2003)}]{meila2003}%
  \BibitemOpen
  \bibfield  {author} {\bibinfo {author} {\bibfnamefont {M.}~\bibnamefont
  {Meil{\u{a}}}},\ }in\ \href@noop {} {\emph {\bibinfo {booktitle} {Learning
  Theory and Kernel Machines}}},\ \bibinfo {editor} {edited by\ \bibinfo
  {editor} {\bibfnamefont {B.}~\bibnamefont {Sch{\"o}lkopf}}\ and\ \bibinfo
  {editor} {\bibfnamefont {M.~K.}\ \bibnamefont {Warmuth}}}\ (\bibinfo
  {publisher} {Springer Berlin Heidelberg},\ \bibinfo {address} {Berlin,
  Heidelberg},\ \bibinfo {year} {2003})\ pp.\ \bibinfo {pages}
  {173--187}\BibitemShut {NoStop}%
\bibitem [{\citenamefont {{Mastrandrea}}\ \emph {et~al.}(2014)\citenamefont
  {{Mastrandrea}}, \citenamefont {{Squartini}}, \citenamefont {{Fagiolo}},\
  and\ \citenamefont {{Garlaschelli}}}]{Mastrandrea2014}%
  \BibitemOpen
  \bibfield  {author} {\bibinfo {author} {\bibfnamefont {R.}~\bibnamefont
  {{Mastrandrea}}}, \bibinfo {author} {\bibfnamefont {T.}~\bibnamefont
  {{Squartini}}}, \bibinfo {author} {\bibfnamefont {G.}~\bibnamefont
  {{Fagiolo}}}, \ and\ \bibinfo {author} {\bibfnamefont {D.}~\bibnamefont
  {{Garlaschelli}}},\ }\href {\doibase 10.1088/1367-2630/16/4/043022}
  {\bibfield  {journal} {\bibinfo  {journal} {New Journal of Physics}\ }\textbf
  {\bibinfo {volume} {16}},\ \bibinfo {pages} {043022} (\bibinfo {year}
  {2014})}\BibitemShut {NoStop}%
\bibitem [{\citenamefont {Sol{\'{e}}-Ribalta}\ \emph
  {et~al.}(2018)\citenamefont {Sol{\'{e}}-Ribalta}, \citenamefont {Tessone},
  \citenamefont {Mariani},\ and\ \citenamefont
  {Borge-Holthoefer}}]{Sole-Ribalta2018}%
  \BibitemOpen
  \bibfield  {author} {\bibinfo {author} {\bibfnamefont {A.}~\bibnamefont
  {Sol{\'{e}}-Ribalta}}, \bibinfo {author} {\bibfnamefont {C.~J.}\ \bibnamefont
  {Tessone}}, \bibinfo {author} {\bibfnamefont {M.~S.}\ \bibnamefont
  {Mariani}}, \ and\ \bibinfo {author} {\bibfnamefont {J.}~\bibnamefont
  {Borge-Holthoefer}},\ }\href {\doibase 10.1103/PhysRevE.97.062302} {\bibfield
   {journal} {\bibinfo  {journal} {Phys. Rev. E}\ } (\bibinfo {year} {2018}),\
  10.1103/PhysRevE.97.062302},\ \Eprint {http://arxiv.org/abs/1801.05620}
  {arXiv:1801.05620} \BibitemShut {NoStop}%
\bibitem [{\citenamefont {Pugliese}\ \emph {et~al.}(2016)\citenamefont
  {Pugliese}, \citenamefont {Zaccaria},\ and\ \citenamefont
  {Pietronero}}]{Pugliese2016a}%
  \BibitemOpen
  \bibfield  {author} {\bibinfo {author} {\bibfnamefont {E.}~\bibnamefont
  {Pugliese}}, \bibinfo {author} {\bibfnamefont {A.}~\bibnamefont {Zaccaria}},
  \ and\ \bibinfo {author} {\bibfnamefont {L.}~\bibnamefont {Pietronero}},\
  }\href {\doibase 10.1140/epjst/e2015-50118-1} {\bibfield  {journal} {\bibinfo
   {journal} {Eur. Phys. J. Spec. Top.}\ }\textbf {\bibinfo {volume} {225}},\
  \bibinfo {pages} {1893} (\bibinfo {year} {2016})},\ \Eprint
  {http://arxiv.org/abs/1410.0249} {arXiv:1410.0249} \BibitemShut {NoStop}%
\end{thebibliography}%

\end{document}